\documentclass[twocolumn,superscriptaddress,floats,prd,nofootinbib,showpacs]{revtex4-2}
\usepackage{anyfontsize} 
\usepackage{stmaryrd}
\usepackage{mathrsfs}
\usepackage{float}
\usepackage{setspace}
\usepackage{amsmath,amssymb,amsfonts}
\usepackage{graphicx}
\usepackage{subfigure}
\usepackage{color} 
\usepackage{fancyhdr}
\usepackage{hyperref} 
\usepackage{enumerate}
\usepackage{CJKutf8}
\newcommand{\Rmnum}[1]{\uppercase\expandafter{\romannumeral #1}}  
\newcommand{\rmnum}[1]{\romannumeral #1}  
\begin{document}
\begin{CJK}{UTF8}{gbsn}

	\title{Higher-dimensional quantum Oppenheimer-Snyder model}
	\author{Zijian Shi}
	\affiliation{Department of Physics, South China University of Technology, Guangzhou 510641, China}
	\author{Xiangdong Zhang}\email{Corresponding author. scxdzhang@scut.edu.cn}
	\affiliation{Department of Physics, South China University of Technology, Guangzhou 510641, China}
	\author{Yongge Ma}\email{Corresponding author. mayg@bnu.edu.cn}
	\affiliation{School of Physics and Astronomy, Key Laboratory of Multiscale Spin Physics, Ministry of Education, Beijing Normal University, Beijing 100875, China}

	\begin{abstract}
		The quantum Oppenheimer-Snyder model for higher-dimensional spacetimes is studied. 
		The higher-dimensional quantum-corrected Schwarzschild black hole is obtained by the junction condition. 
		It turns out that quantum bounces always occur in the collapse thus that the classical gravitational collapse singularities are avoided. 
		The scalar perturbations upon the quantum-corrected black holes are also studied. 
		It turns out that the quantum corrections enhance the oscillation frequency in lower dimensions and decrease it in higher dimensions. 
		Moreover, the thermodynamic laws of the quantum-corrected black holes 
		imply that the Hawking temperature of quantum-corrected black hole decreases as the mass decreases in contrast to the classical situation. 
		The behaviour of heat capacity indicates that quantum corrections introduce an extra phase transition of the black holes. 
	\end{abstract}

	\maketitle

	\section{Introduction}\label{Intro}
	
	As one of the most fascinating predictions of Einstein's general theory of relativity, 
	black hole (BH) has intrigued and puzzled physicists for decades. 
	There is accumulating observational evidence supporting the existence of these enigmatic objects, 
	including direct observations such as gravitational waves and images from the 
	Event Horizon Telescope \cite{EventHorizonTelescope:2019dse,EventHorizonTelescope:2022wkp,LIGOScientific:2016aoc}, 
	while many fundamental issues on BH remain unanswered, including the singularity inside BH \cite{Cardoso:2019rvt}. 
	According to the singularity theorems by Penrose and Hawking \cite{Penrose:1964wq,Hawking:1970zqf}, under quite general conditions, 
	singularities are unavoidable, implying the breakdown of the classical theory in these extremely curved 
	regions of spacetime. It is widely believed that a quantum theory of gravity, 
	which unifies quantum mechanics and general relativity, should be set up to resolve the singularities. 
	Among the different candidate quantum gravity theories, one promising approach is loop quantum gravity (LQG), 
	which provides a fascinating perspective on the quantum structure 
	of spacetime and offers potential insights into the issue of singularities \cite{Rovelli:1997yv,Thiemann:2001gmi,Ashtekar:2004eh,Han:2005km}. 
	In recent years, significant progress has been made in LQG, 
	particularly for understanding the quantum properties of BH \cite{Ashtekar:1997yu,Ashtekar:2005qt,Rovelli:1996dv,Zhang:2023yps,Perez:2017cmj}. 

	In classical general relativity, the Oppenheimer-Snyder (OS) model is widely adopted to
	describe the gravitational collapse of a massive star \cite{Oppenheimer:1939ue}. 
	It assumes that the star is spherically symmetric and the collapse is homogenous. 
	Although the OS model is simple, it is able to capture the basic features of stellar collapse. 
	One important result of the OS model is that the mass of a BH cannot be less than 
	certain critical value, known as the Tolman–Oppenheimer–Volkoff (TOV) limit \cite{Tolman:1939jz} 
	which is equal to about three solar masses. 
	The OS model also predicts that the density of mass would become infinite as the singularity is formed. 
	Recently, by combining the OS model with LQG model, 
	a four-dimensional nonsingular extension of Schwarzschild BH was found \cite{Lewandowski:2022zce}. 
	The resulting spacetime configuration contains a BH to white hole transition \cite{Han:2023wxg}. 
	A similar idea is also used to obtain a regular BH spacetime from an effective Lagrangian for quantum Einstein gravity \cite{Bonanno24}. 

	Higher-dimensional spacetime is a vibrant and challenging area of research. 
	One of the primary motivations for studying higher-dimensional theories is the potential to unify 
	gravity with other fundamental interactions \cite{Horava:1995qa}. The standard model of particle physics, 
	as our current best theory of fundamental interactions, does not include gravity. 
	The idea for higher-dimensional theories (e.g., Kaluza-Klein theory \cite{Kaluza:1921tu,Klein:1926tv,Bailin:1987jd} and M-theory \cite{Witten:1995em}) 
	is that matter fields are the consequence of extra spatial dimensions, 
	offering a unified framework for all fundamental interactions. 
	Higher-dimensional theories can also shed light on some of the most profound mysteries of the universe, 
	for instance, the nature of dark matter and dark energy \cite{Dvali:2000hr,Qiang:2004gg,Zhang:2012em}. 
	While all the current studies on loop quantum BH models are limited to the four-dimensional case, 
	the aim of this paper is to generalize the four-dimensional quantum OS model to higher-dimensional case to 
	obtain a nonsingular higher-dimensional quantum Schwarzschild BH model and study its properties. 

	A key question about BH spacetimes is whether they remain stable under perturbations \cite{Regge:1957td}. 
	The perturbations of BH can be divided into two categories: exogenous perturbations and endogenous perturbations. 
	The former refers to the influence of matter or energy from outside the BH on the spacetime, 
	such as gravitational perturbations caused by a star or planet orbiting the BH. 
	The latter refers to perturbations from the BH itself, 
	such as those generated by BH's mass, spin or electric charge. 
	The response of BH to these perturbations can be divided into three stages: 
	Initial Burst, Quasinormal Modes (QNMs) and Late Tails. 
	QNMs represent the phenomenon that the perturbations 
	oscillate at a specific frequency and gradually decay away. 
	This behavior is only dependent on the spacetime itself and has nothing to do with 
	the initial perturbations. 
	Therefore, it is called the characteristic sound of the BH \cite{Leaver:1986gd,Berti:2009kk}. 

	In 1973, Bekenstein established a correspondence between the parameters of a stationary BH and 
	traditional thermodynamic parameters, proposing the laws of BH thermodynamics \cite{Bekenstein:1973ur}. 
	In 1975, Hawking applied quantum field theory to BH spacetime and discovered the thermal radiation \cite{Hawking:1975vcx}. 
	This discovery laid a solid foundation for BH thermodynamics, 
	showing that a BH is a genuine thermodynamic system. 
	The four laws of BH thermodynamics were formulated \cite{Bardeen:1973gs}. 
	Since then, extensive research has been conducted to further understand the fundamental properties of BH
	through its thermodynamics \cite{Hawking:1982dh,Wald:1993nt}. 
	Hence, in this paper, we will also study the QNMs and the thermodynamic properties
	of the higher-dimensional quantum Schwarzschild BH model. 

	The main content of this paper is as follows: 
	Firstly, we derive the metric of the higher-dimensional quantum-corrected BH in Sec. \ref{model} and study its physical properties. 
	In Sec. \ref{QNMs}, we investigate the QNMs generated by perturbations of the higher-dimensional BH. 
	In Sec. \ref{thermodynamics}, we will study the thermodynamic properties of the higher-dimensional quantum-corrected BH. 
	Finally, our results will be discussed and summarized in Sec. \ref{conclusion}.

	\section{Quantum Oppenheimer-Snyder model in higher dimensions}\label{model}

	\subsection{Quantum-corrected Metric}\label{spacetime}
	In the four-dimensional quantum Oppenheimer-Snyder model proposed in \cite{Lewandowski:2022zce,Yang:2022btw}, 
	One starts from an effective interior spacetime region that incorporates the correction of loop quantum cosmology (LQC) \cite{Ashtekar:2006wn}.
	The effective interior spacetime region is then matched to an exterior spacetime region via junction conditions 
	on their common boundary surface. Now, we will apply this method to an arbitrary $(d+1)$-dimensional spacetime.
	
	We consider the gravitational collapse of a spherically symmetric, homogeneous and pressureless star 
	in $(d+1)$-dimensional asymptotically flat spacetime. Due to the homogeneity and isotropy, 
	the spacetime region inside the star can be described by the Friedmann-Robertson-Walker (FRW) metric as in cosmology,  
	\begin{equation}\label{FRW}
		\mathrm{d}s_{in}^2=-\mathrm{d}T^2+a^2(T)[\mathrm{d}R^2+R^2\mathrm{d}\Omega^2] ,
	\end{equation}
	where $a(T)$ is the scale factor and $\mathrm{d}\Omega^2$ is the metric of the $(d-1)$-sphere which reads
	\begin{eqnarray}
		\mathrm{d}\Omega^2=
		&&\mathrm{d}\theta_1^2+sin^2\theta_1\mathrm{d}\theta_2^2+......    \notag \\
		&&+sin^2\theta_1sin^2\theta_2...sin^2\theta_{d-2}\mathrm{d}\theta_{d-1}^2 .
	\end{eqnarray}
	The range of $\theta_2$ is $\left[0,2\pi\right)$, while those of the other angles are $\left[0,\pi\right]$. 
	All the dynamical information of the FRW solution is encoded in the scale factor $a(T)$. 
	Let the radial coordinate of the surface of the star be $R_0$. 
	Then the physical radius of the star reads $\tilde{R}=a(T)R_0$. 
	In the classical theory, the dynamics of the star's interior is governed by the 
	$(d+1)$-dimensional Friedmann equation
	\begin{equation}\label{CFE}
		H^2=\frac{2\kappa}{d(d-1)}\rho ,
	\end{equation}
	where
	\begin{equation}
		H=\frac{\dot{a}}a
	\end{equation}
	is the Hubble parameter related to the scale factor and its derivative $\dot{a}$ with respect to $T$. 
	The constant $\kappa$ is related to the Newtonian gravitational constant $G$ by $\kappa=8\pi G$. 
	The matter density $\rho$ in $(d+1)$-dimensional spacetime satisfies $\rho=\frac{M}{V}$, 
	where $M$ and $V$ is the mass and volume of the star.
	The stellar volume $V$ is simply the volume of a $d$-dimensional ball with radius $\tilde{R}$ as
	\begin{equation}\label{Volume}
		V=\frac{8\pi \tilde{R}^d}{d(d-1)\mu} ,
	\end{equation}
	with 
	\begin{equation}
		\mu=\frac{8\pi}{(d-1)\Omega} ,
	\end{equation}
	and 
	\begin{equation}
		\Omega=\frac{d\pi^{d/2}}{\Gamma\left(\frac{d}{2}+1\right)} ,
	\end{equation}
	where $\Omega$ is the area of the $(d-1)$-dimensional unit sphere and 
	$\Gamma$ is the Gamma function \cite{Fernandes:2023byx}.
	Therefore, by Eq. \eqref{Volume}, Eq. \eqref{CFE} can be rewritten as 
	\begin{equation}\label{cFriedman}
		H^2=\frac{2GM\mu}{\tilde{R}^{d}} .
	\end{equation}
	Recall that the higher-dimensional LQC has already been well established in \cite{Zhang:2015bxa}. 
	The resulting effective Friedman equation takes the form of 
	\begin{equation}\label{QFE}
		H^2=\frac{2\kappa}{d(d-1)}\rho\left(1-\frac\rho{\rho_c}\right) ,
	\end{equation}
	where 
	\begin{equation}\label{rhoc}
		\begin{aligned}\rho_c&=\frac{d(d-1)}{2\kappa\gamma^2(\Delta)^{\frac{2}{d-1}}}\end{aligned}
	\end{equation}
	represents the critical matter density in $(d+1)$-dimensional spacetime. 
	Here $\gamma$ is a nonzero real number corresponding to the Immirzi parameter in four dimensions, and 
	$\begin{aligned}\Delta=8\sqrt{d}\pi\gamma(\ell_{\mathrm{p}})^{d-1}\end{aligned}$ 
	is the area gap, with $\ell_{\mathrm{p}}=\sqrt[d-1]{G\hbar}$ being the Planck length \cite{Bodendorfer:2011nx}. 
	Similar to the classical case, Eq. \eqref{QFE} can also be expressed in the following form, 
	\begin{equation}\label{LQCFriedman}
		H^2=\frac{2GM\mu}{\tilde{R}^{d}}-\frac{4{G}^2{M}^2\gamma^2\mu^2\Delta^{\frac2{d-1}}}{\tilde{R}^{2d}} .
	\end{equation}
	The presence of the last term evolving $\rho_c$ in Eq. \eqref{QFE} indicates that 
	during the collapsing procedure of the star, its density can reach $\rho_c$, 
	so that the collapse would turn into a bounce, and hence the singularity is avoided.

	The exterior of the star is assumed to be a static and spherically symmetric spacetime region, 
	and hence the metric can be expressed in a general form using Schwarzschild coordinates, 
	\begin{equation}\label{EM0}
		\mathrm{d}s_{out}^2=-f(r)\mathrm{d}t^2+g(r)^{-1}\mathrm{d}r^2+r^2\mathrm{d}\Omega^2 .
	\end{equation}
	We then impose the continuity of the metric and extrinsic curvature on the interface between 
	the star and the exterior vacuum, which is commonly referred as the Darmois-Israel junction conditions \cite{Darmois:1927,Israel:1966rt}. 
	By applying the junction conditions, the functions $f(r)$ and $g(r)$ in Eq. \eqref{EM0} can be determined. 
	We leave the detailed derivation in the appendix A. 

	By using the classical and quantum-corrected forms of Friedman equations \eqref{cFriedman} and \eqref{LQCFriedman}, 
	the metrics of the external spacetime region are derived respectively as 
	\begin{equation}\label{CES}
		\mathrm{d}s_{c}^2=-(1-\frac{2GM\mu}{r^{d-2}})\mathrm{d}t^2+(1-\frac{2GM\mu}{r^{d-2}})^{-1}\mathrm{d}r^2+r^2\mathrm{d}\Omega^2 
	\end{equation}
	and 
	\begin{eqnarray}\label{QES}
		\mathrm{d}s_{q}^2=
		&&-(1-\frac{2GM\mu}{r^{d-2}}+\frac{4{G}^2{M}^2\mu^2\gamma^2\Delta^{\frac2{d-1}}}{r^{2d-2}})\mathrm{d}t^2 \notag \\
		&&+(1-\frac{2GM\mu}{r^{d-2}}+\frac{4{G}^2{M}^2\mu^2\gamma^2\Delta^{\frac2{d-1}}}{r^{2d-2}})^{-1}\mathrm{d}r^2 \notag \\
		&&+r^2\mathrm{d}\Omega^2 .
	\end{eqnarray}
	Note that Eq. \eqref{CES} is just the $(d+1)$-dimensional Schwarzschild metric, 
	while Eq. \eqref{QES} is the corresponding quantum deformation of $(d+1)$-dimensional Schwarzschild metric. 
	Note also that the mass $M$ of the star appears in Eq. \eqref{CES} and Eq. \eqref{QES} 
	coincides with the ADM mass.

	\subsection{Bounce and Geodesics}\label{geodesics}
	We now delve into the quantum-corrected spacetime described by Eq. \eqref{QES} and to explore its physical properties. 
	As shown in appendix A, the physical radius of the star satisfies $\dot{r}^2=1-f$. 
	Therefore, when the radius reaches the point $r=\left(2GM\mu\alpha\right)^{\frac1d} \equiv r_b $ that $1-f=0$, 
	where $\alpha\equiv\gamma^2\Delta^{\frac2{d-1}}$, one gets $\dot{r}=0$. 
	This indicates that the collapse of the star halts, 
	reaching the minimum radius $r_b$ and maximum energy density $\rho_c$.
	Subsequently, the star will shift into an expanding state, indicating a bounce.
	In this way, the radius of the star will not decrease to zero and thus the classical singularity is avoided. 
	Since Eq. \eqref{QES} describes the external spacetime of the star, it is well-defined for 
	$\left.r\in\left[\begin{matrix}r_b,\infty\end{matrix}\right.\right)$. 
	This is of course different from the classical higher dimensional Schwarzschild case 
	as in Eq. \eqref{CES} where $r=0$ is a singularity.

	We are interested in exploring the overall configuration of the metric \eqref{QES}, 
	which is contingent upon the number of solutions for $f(r)=0$. 
	As shown in appendix B, the number of solutions depends on the stellar mass $M$. 
	There is a lower mass bound $M_c$ for the existence of a solution. 
	There are following three different cases. 
	\begin{enumerate}[i)]
		\item $M>M_c$, $f(r)=0$ has two solutions $r=r_{\pm}$.
		They are the two horizons of the spacetime. 
		Since the normal vectors $\partial/\partial t$ of $r=r_{\pm}$ belong to a Killing vector field, which satisfies 
		$\mathrm{g}_{ab}^{out}\left(\frac{\partial}{\partial t}\right)^a\left(\frac{\partial}{\partial t}\right)^b=-f(r)=0$ 
		on the two horizons, $r=r_{\pm}$ are Killing horizons. 
		A detailed analysis shows that: the outer horizon $r_+$ is an event horizon, 
		while the inner horizon $r_-$ is a Cauchy horizon. 
		For instance, the Penrose diagram of the maximally extended spacetime for metric \eqref{QES} 
		in five-dimensional case is shown in Fig. \ref{Penrose-diagram}. 
		The maximally extended spacetime consists of three types of regions as shown in Table \ref{regions}. 
		The red line in Fig. \ref{Penrose-diagram} represents the worldline of the surface of the collapsing star. 
		During the collapsing process, the star first shrinks, and then bounces once the radius reaches $r_b$. 
		Afterwards, the star will begin to expand and successively appear in \Rmnum{3'} and \Rmnum{1'} regions. 
		Although the collapse does not form a singularity, it does form an event horizon. 
		Therefore, in the asymptotically flat region \Rmnum{1}, a BH is formed. 
		In the asymptotically flat \Rmnum{1'} region, a white hole is formed due to the expansion of the star. 
		The \Rmnum{5} region plays a role of a tunnel connecting regions \Rmnum{2} and \Rmnum{3'}. 
		Thus, the collapse of a star would lead to a phenomenon of a BH to white hole transition. 
		However, the resulting white hole exists in a future universe. 
		It is worth noting that the region to the left of the red line corresponds to the interior of the star and 
		should therefore be characterized by the metric described by Eqs. \eqref{FRW} and \eqref{QFE}.
		\item $M=M_c$, $f(r)=0$ has exactly one solution.
		In this scenario, the two horizons $r_-$ and $r_+$ overlap to form a single event horizon, 
		leading to the disappearance of the region $r_{-}<r<r_{+}$. 
		Nevertheless, the phenomena witnessed by observers in regions \Rmnum{1} and \Rmnum{1'} will be 
		similar to those of the case $M>M_c$.
		\item $M<M_c$, $f(r)=0$ has no solution.
		In this scenario, the star cannot form an event horizon even as it collapses to the minimum radius $r_b$, 
		where the star undergoes a bounce. 
		The global causal structure of the maximally extended spacetime for the exterior metric is 
		identical to that of Minkowski spacetime.
		\begin{table}[!htb]
			\caption{In case (\rmnum{1}), the spacetime consists of three types of regions.}
			\begin{ruledtabular}
				\begin{tabular}{lll}
					region            & example                                       & characteristic               \\
					\hline
					$r_{+}<r<\infty$  & \Rmnum{1}, \Rmnum{1'}, \Rmnum{4}, \Rmnum{4'}  & asymptotically flat, static  \\
					$r_{-}<r<r_{+}$   & \Rmnum{2}, \Rmnum{2'}, \Rmnum{3}, \Rmnum{3'}  & dynamic                      \\
					$r_{b}<r<r_{-}$   & \Rmnum{5}, \Rmnum{6}                          & static                       \\
				\end{tabular}
			\end{ruledtabular}
			\label{regions}
		\end{table}
		\begin{figure}[!htb]
			\includegraphics [width=0.45\textwidth]{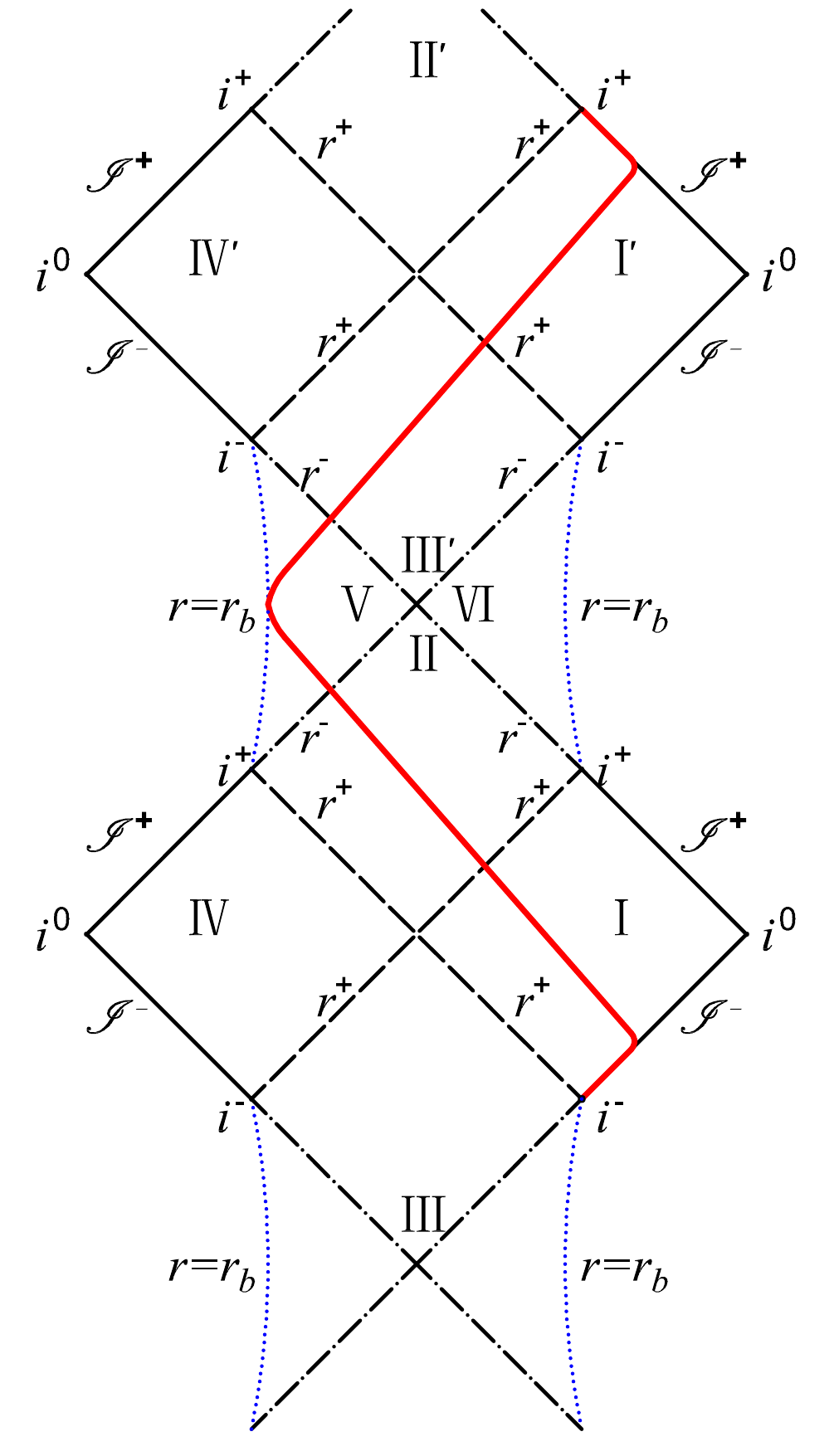}
			\caption{The Penrose diagram of a maximally extended quantum-corrected black hole for $d=4$. 
			The blue dashed line corresponds to the minimum radius of the star, $r=r_b$. 
			The red line represents the worldline of the stellar surface. 
			}
			\label{Penrose-diagram}
		\end{figure}
	\end{enumerate}

	\section{Quasinormal Modes}\label{QNMs}

	\subsection{Black Hole Perturbations and the Wave-like Equation}\label{BHP}
	To understand the spacetime properties of metric \eqref{QES}, 
	we focus on the perturbations of a massless scalar field in the quantum-corrected BH. 
	The dynamics of the massless scalar field $\psi$ in curved spacetime can be 
	described by the Klein-Gordon (KG) equation 
	\begin{equation}
		\begin{aligned}\frac{1}{\sqrt{-\mathrm{g}}}\partial_{\mu}\left(\sqrt{-\mathrm{g}}\mathrm{g}^{\mu\nu}\partial_{\nu}\psi\right)=0.\end{aligned}
	\end{equation}
	As the spacetime outside the horizon is static and spherically symmetric, 
	we can perform the following separation of variables for $\psi$, 
	\begin{equation}\label{SOV}
		\begin{aligned}\psi(t,r,\theta_i)=\sum_{l,m}e^{-i\omega t}r^{\frac{1-d}2}\Phi_l(r)Y_{lm}(\theta_i),\end{aligned}
	\end{equation}
	where $Y_{lm}(\theta_i)$ are the spherical harmonics of $d-1$ degrees, 
	with $l$ and $m$ being the multipole and azimuthal quantum number, respectively. 
	We substitute Eqs. \eqref{SOV} and \eqref{QES} into the KG equation to obtain the radial wave equation: 
	\begin{eqnarray}\label{radial-wave-equation}
		&&f^2\Phi_l^{\prime\prime}+ff^{\prime}\Phi_l^{\prime}+\omega^2\Phi_l- \notag \\
		&&\left[\frac{l(l+d-2)}{r^2}+\frac{(d-1)(d-3)}{4r^2}f+\frac{d-1}{2r}f^{\prime}\right]f\Phi_l \notag \\
		&&=0,
	\end{eqnarray} 
	where $f'\equiv\mathrm{d}f(r)/\mathrm{d}r$, $\Phi_l^{\prime}\equiv\mathrm{d}\Phi_l(r)/\mathrm{d}r$ and 
	$\Phi_l^{\prime\prime}\equiv\mathrm{d}^2\Phi_l(r)/\mathrm{d}r^2$ \cite{Guo:2020caw}.
	In the tortoise coordinate, 
	$\mathrm{d}r_*=\mathrm{d}r/f(r)$, Eq. \eqref{radial-wave-equation} takes the following form: 
	\begin{equation}\label{time-independent}
		\begin{aligned}\frac{\mathrm{d}^2\Phi_l(r_*)}{\mathrm{d}r_*^2}+[\omega^2-V(r)]\Phi_l(r_*)=0\end{aligned}
	\end{equation}
	where 
	\begin{equation}
		V(r)=\left[\frac{l(l+d-2)}{r^2}+\frac{(d-1)(d-3)}{4r^2}f+\frac{d-1}{2r}f^{\prime}\right]f
	\end{equation}
	is the effective potential \cite{Cai:2020igv,Natario:2004jd}.

	In order to determine the eigen-frequency $\omega$, 
	it is necessary to impose appropriate boundary conditions on Eq. \eqref{time-independent}. 
	For the event horizon, $r\rightarrow r_+$ leads to $r_*\rightarrow-\infty$ and $V\rightarrow0$. 
	Classically, nothing can escape from the event horizon, so only incoming waves exist. 
	The asymptotic solution in this case is $\Phi_l(r_*)\sim e^{-i\omega r_*}$. 
	Since the metric is asymptotically flat, we consider 
	$r\rightarrow+\infty$, and it leads to $r_*\rightarrow+\infty$ and $V\rightarrow0$. 
	Physically, there are no incident waves coming from infinity. 
	Therefore, the asymptotic solution in this case is $\Phi_l(r_*)\sim e^{+i\omega r_*}$.

	If we consider the time-dependent case and denote part of Eq. \eqref{SOV} as $\Psi=e^{-i\omega t}\Phi_l(r)$, 
	the wave-like equation becomes 
	\begin{equation}\label{time-dependent}
		-\frac{\partial^2\Psi}{\partial t^2}+\frac{\partial^2\Psi}{\partial r_*^2}-V(r)\Psi=0 .
	\end{equation}

	\subsection{Numerical Methods}\label{Numerical-Methods}
	In order to calculate QNMs, we will introduce the WKB approximation and Finite Element Method (FEM) in this subsection. 

	The first-order WKB technique was first employed to investigate the 
	scattering problem around black holes by Schutz and Will \cite{Schutz:1985}.
	This semi-analytic method is applicable to effective potentials characterized by a barrier structure, 
	which exhibit constant values both at the event horizon and spatial infinity. 
	The wave function is expanded as a series at the two asymptotic regions and it would be matched with 
	the Taylor expansion near the peak of the effective potential through the two turning points. 
	Following that, Iyer and Will extended this technique to third WKB order, 
	surpassing the limitation of the eikonal approximation \cite{Iyer:1986np}.
	Then, Konoplya made significant progress by extending it to the sixth order \cite{Konoplya:2003ii}.
	Recently, Matyjasek and Opala introduced the method of Pade approximation, 
	which enhanced the precision of the calculation to the 13th order \cite{Matyjasek:2017psv}.
	The frequency resulted from the WKB approximation can be written in the following general form \cite{Konoplya:2019hlu}, 
	\begin{eqnarray} \label{WKBFun}
		\omega^{2} 
		&=&
		V_{0}+A_{2}\left(\mathcal{K}^{2}\right)+A_{4}\left(\mathcal{K}^{2}\right)+A_{6}\left(\mathcal{K}^{2}\right)+\ldots \notag  \\ 
		&& -\mathrm{i} \mathcal{K} \sqrt{-2 V_{2}}\left(1+A_{3}\left(\mathcal{K}^{2}\right)+A_{5}\left(\mathcal{K}^{2}\right)+\ldots\right),
	\end{eqnarray}
	where 
	\begin{equation}
		\mathcal{K}=n+\frac{1}{2},\quad n=0,1,2,\ldots,
	\end{equation}
	with $n$ representing the overtone number, and
	$V_{i},i=0, 2,3...$ is the value of $V(r)$ and its $i$th derivative at the maximum of effective potential. 
	The $k$th order correction term $A_{k}\left(\mathcal{K}^{2}\right),k=2,3...$ is a polynomial of 
	$V_2$,$V_3$,$\ldots$,$V_{2k}$ and $\mathcal{K}^{2}$ with rational coefficients. 
	It is worth noting that the WKB formula typically exhibits superior accuracy when $l>n$. 
	However, when $l\leq n$, this approach does not always yield dependable results \cite{Konoplya:2003ii,Konoplya:2019hlu}. 
	We will utilize the 13th-order WKB method with Pade approximants to 
	calculate the fundamental mode ($n=0$), which is the slowest decaying mode.

	The idea of the FEM is to replace continuous differentials with a series of discrete differences \cite{Fu:2022cul,Liu:2023kxd}. 
	Therefore, instead of directly solving Eq. \eqref{time-dependent}, an equivalent approach is to handle the following equation, 
	\begin{eqnarray}
		\Psi_{j}^{i+1}=&&-\Psi_j^{i-1}+\left(2-2\frac{\delta t^2}{\delta r_*^2}-\delta t^2V_j\right)\Psi_j^i \notag \\
					   &&+\frac{\delta t^2}{\delta r_*^2}\left(\Psi_{j-1}^i+\Psi_{j+1}^i\right),
	\end{eqnarray}
	where $t_i=t_0+i\delta t$, $r_{*j}=r_{*0}+j\delta r_*$, $\Psi_j^i=\Psi\left(t=t_i,r_*=r_{*j}\right)$ and $V_j=V\left(r_*=r_{*j}\right)$.
	The initial conditions can be set to be 
	\begin{equation}
		\Psi\left(r_*,t_0\right)=C_1\exp\left(-C_2\left(r_*-C_3\right)^2\right) ,
	\end{equation}
	\begin{equation}
		\frac{\partial}{\partial t}\Psi\left.(r_*,t)\right|_{t=t_0}=0 .
	\end{equation}
	We choose $\delta t/\delta r_{*}=2/3$ and $C_1=10$, $C_2=1/8$, $C_3=r_*(V_{max})+30$ in our calculation \cite{Liu:2023kxd}.

	\subsection{QNMs Results}\label{QNMsresult}

	In this subsection, we will discuss the results of the quasinormal modes. 
	We consider the multi-effects of the following factors on QNMs: 
	stellar mass, spacetime dimensions, and quantum corrections. 
	It should be noted that the real and imaginary parts of QNMs represent 
	the actual oscillation frequencies and the damping rates of the perturbation field, respectively. 
	In the calculations, it has been observed that the imaginary part is always negative, 
	hence we will use the absolute value $|\operatorname{Im(\omega)}|$. 
	We set some of parameters as $\left\{G\to1, \alpha\to1\right\}$.

	\subsubsection{Influence of the stellar mass} 
\begin{figure*}[!htbp]
		\includegraphics [width=1.0\textwidth]{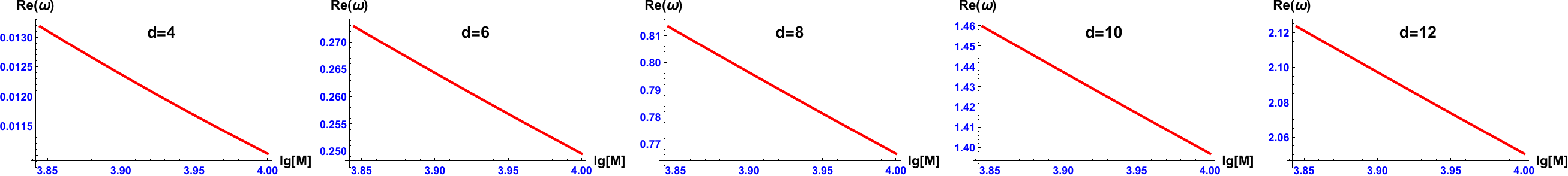}
		\caption{The real part of QNMs with respect to the mass of the quantum-corrected black hole when $n=0$, $l=1$.}
		\label{Re-vs-M}
	\end{figure*}

\begin{figure*}[!htbp]
		\includegraphics [width=1.0\textwidth]{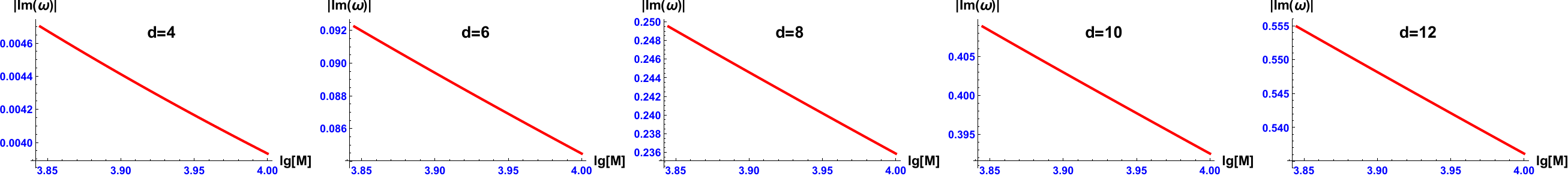}
		\caption{The absolute value of the imaginary part of QNMs with respect to the mass of the quantum-corrected black hole when $n=0$, $l=1$.}
		\label{Im-vs-M}
	\end{figure*}

	We set $l=1$ and calculate the quasinormal mode frequencies for spacetimes of different dimensions 
	with stellar masses $M$ ranging from $7000$ to $10000$. 
	We plot the real parts and absolute value of the imaginary parts 
	with respect to $M$ in Fig. \ref{Re-vs-M} and Fig. \ref{Im-vs-M}. 
	From those figures, we observe that an increase in stellar mass leads to a monotonic decrease in 
	both the real part and absolute value of the imaginary part, indicating that higher stellar mass results in 
	lower oscillation frequency and damping rate of the scalar field in the quantum-corrected black hole.

	\subsubsection{Influence of the spacetime dimensions}
\begin{figure}[!htbp]
		\includegraphics [width=0.48\textwidth]{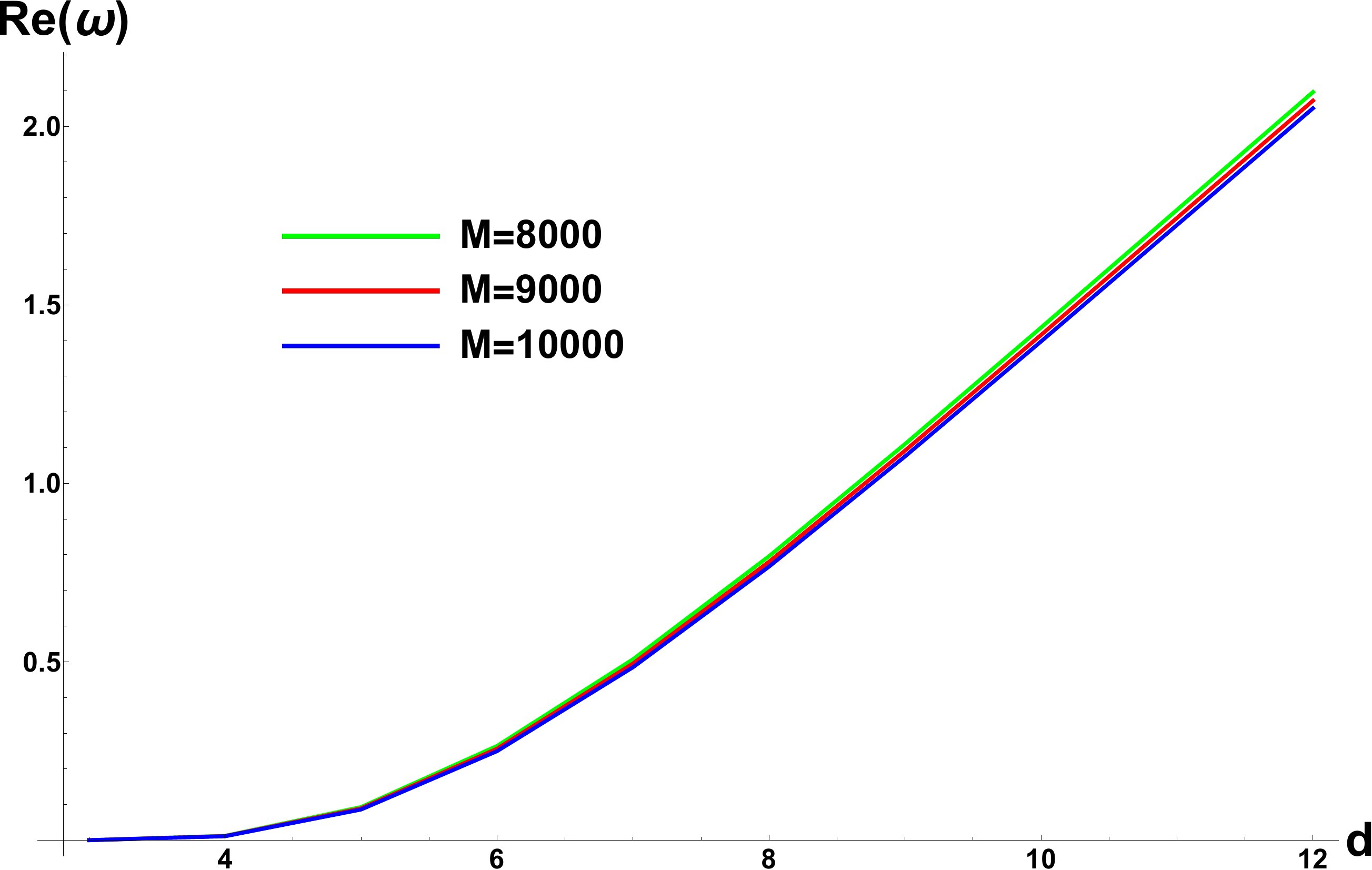}
		\caption{The real part of the QNMs with respect to the dimensions when $n=0$, $l=1$.}
		\label{Re-vs-d}
	\end{figure}

\begin{figure}[!htbp]
		\includegraphics [width=0.48\textwidth]{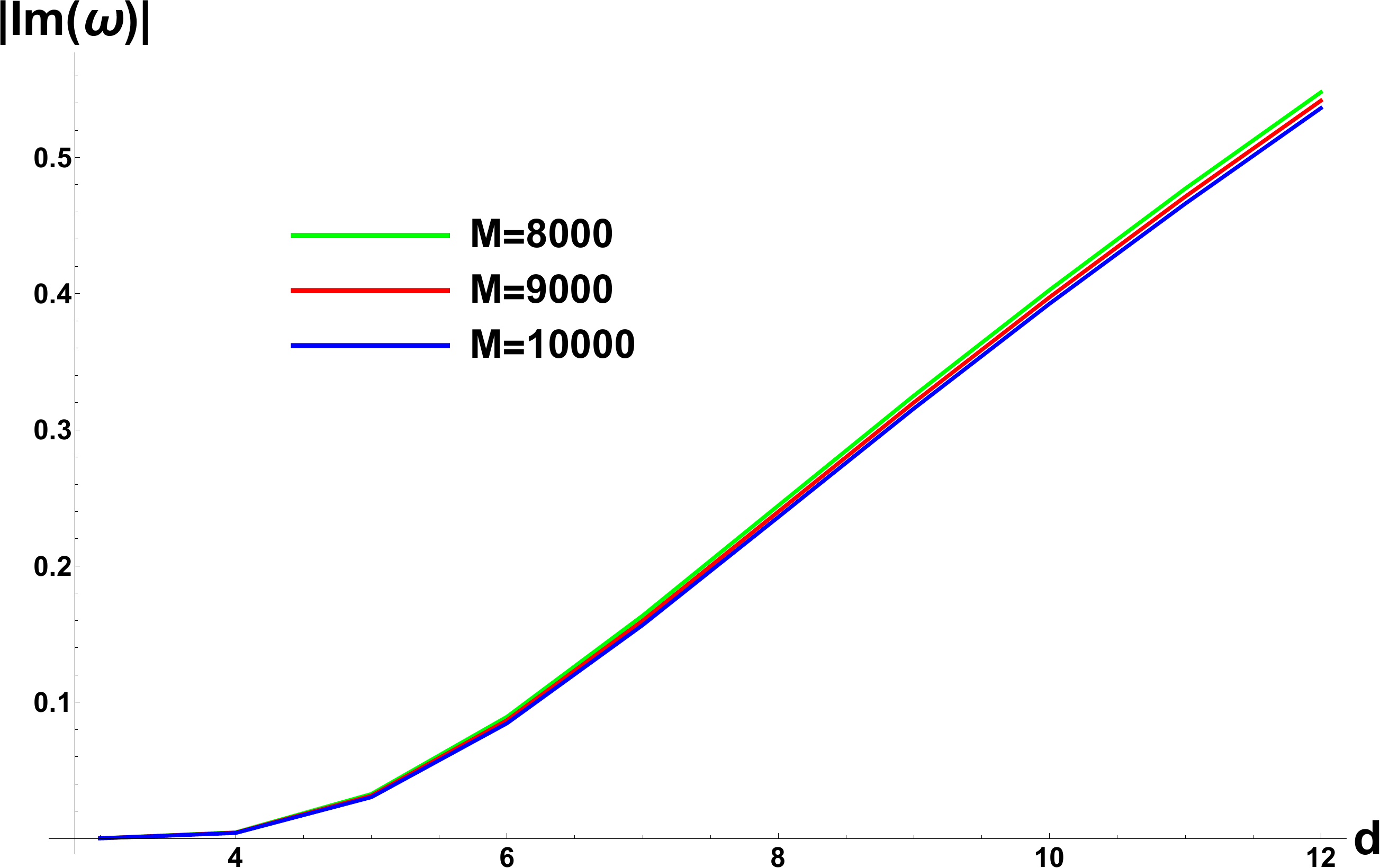}
		\caption{The absolute value of the imaginary part of the QNMs with respect to the dimensions when $n=0$, $l=1$.}
		\label{Im-vs-d}
	\end{figure}

	We depict the variation of QNMs with spacetime dimensions 
	for $M=8000$, $9000$, and $10000$ in Fig. \ref{Re-vs-d} and Fig. \ref{Im-vs-d}. 
	We observe that the increase in spacetime dimension leads to a monotonic increase in 
	both the real part and absolute value of the imaginary part for a fixed $M$. 
	It means that for a star of a given mass, increasing the dimensionality of spacetime leads to the increase of the 
	oscillation frequencies and damping rates for the scalar field.

	\subsubsection{Influence of the quantum corrections}
\begin{figure*}[!htbp]
		\includegraphics [width=1.0\textwidth]{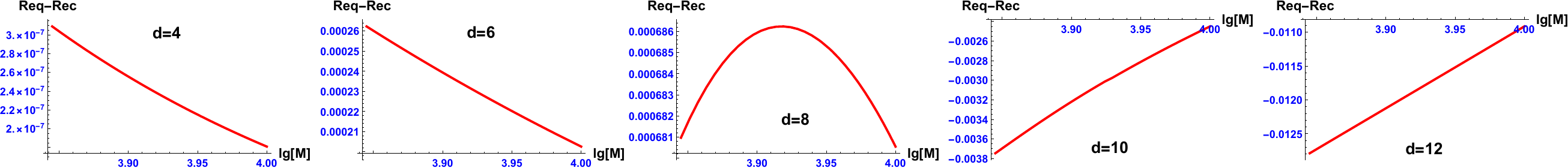}
		\caption{$Req$ is the real part of the QNMs after quantum corrections, 
		while $Rec$ is the real part of QNMs in the classical (Schwarzschild) case. 
		The figure shows how their difference varies with the black hole mass when $n=0$, $l=1$.}
		\label{DRe-vs-M}
	\end{figure*}

\begin{figure*}[!htbp]
		\includegraphics [width=1.0\textwidth]{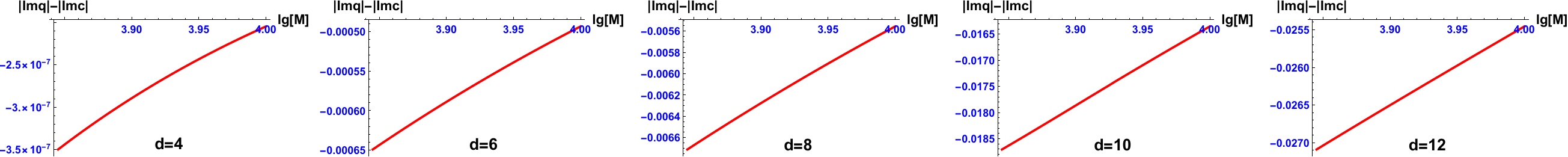}
		\caption{$|Imq|$ is absolute value of the imaginary part of the QNMs after quantum corrections, 
		while $|Imc|$ is the value in the classical (Schwarzschild) case. 
		The figure shows how their difference varies with the black hole mass when $n=0$, $l=1$.}
		\label{DIm-vs-M}
	\end{figure*}

	We calculate the differences in QNMs under quantum corrections and classical Schwarzschild cases. 
	We plot the differences with respect to $M$ in Fig. \ref{DRe-vs-M} and Fig. \ref{DIm-vs-M}.

	We observe that quantum corrections always result in a decrease in the absolute value of the imaginary part. 
	However, the larger the mass of the BH, the weaker this correction to the imaginary part becomes. 
	The situation is different for the real part. 
	In low-dimensional spacetimes, quantum corrections always lead to an increase in the real part, 
	with this discrepancy diminishing as the mass increases. 
	Conversely, in high-dimensional spacetimes, quantum corrections always result in a decrease in the real part, 
	with the discrepancy also decreasing as mass increases. 
	However, transitional scenarios exist between the two cases, 
	such as the nine-dimensional spacetime ($d=8$) depicted in Fig. \ref{DRe-vs-M}. 



	
\begin{figure}[!htbp]
		\includegraphics [width=0.48\textwidth]{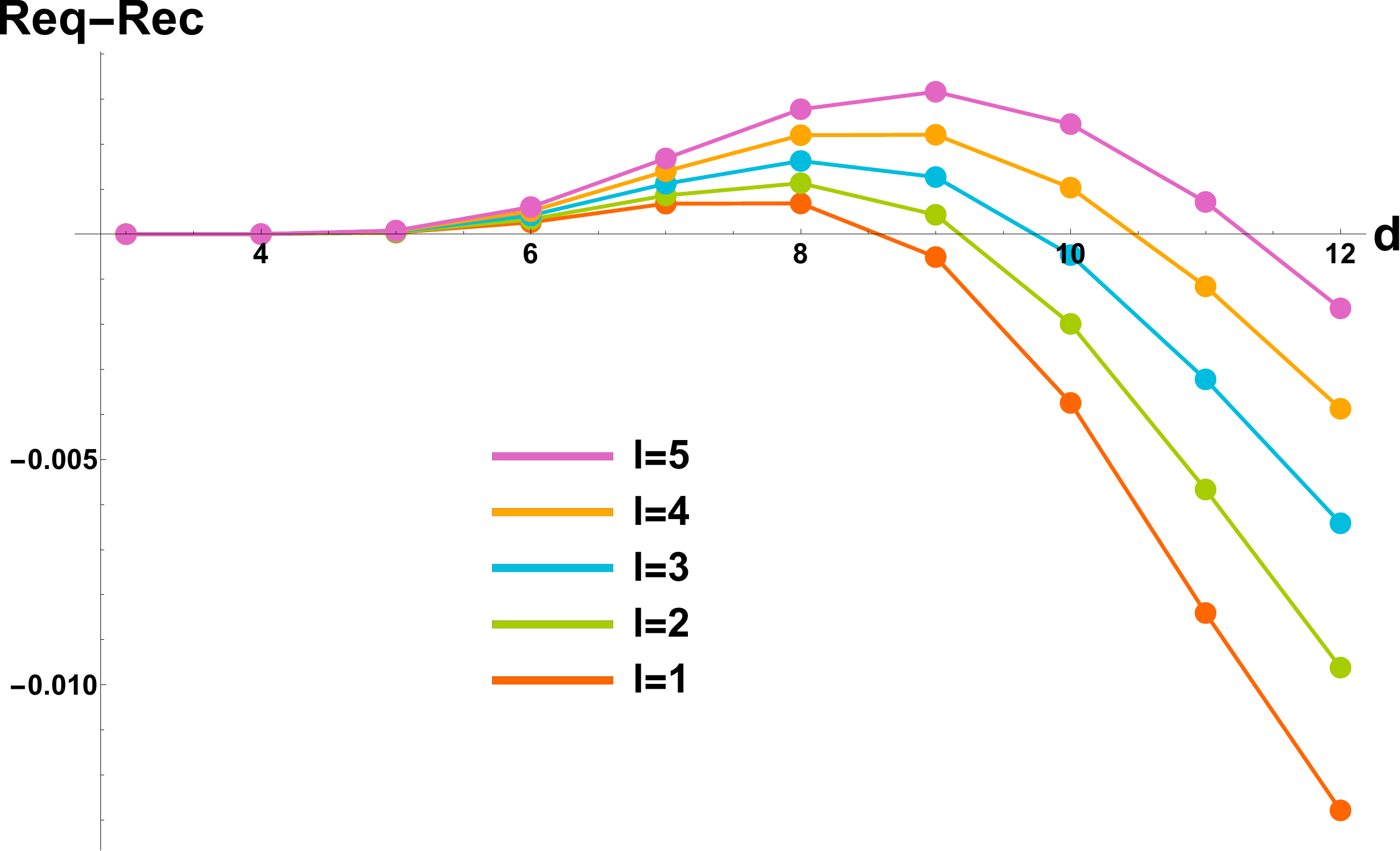}
		\caption{The difference in the real parts of QNMs with respect to the dimensions of spacetime when $M=7000$, $n=0$.}
		\label{DRe-vs-d}
	\end{figure}

\begin{figure}[!htbp]
		\includegraphics [width=0.48\textwidth]{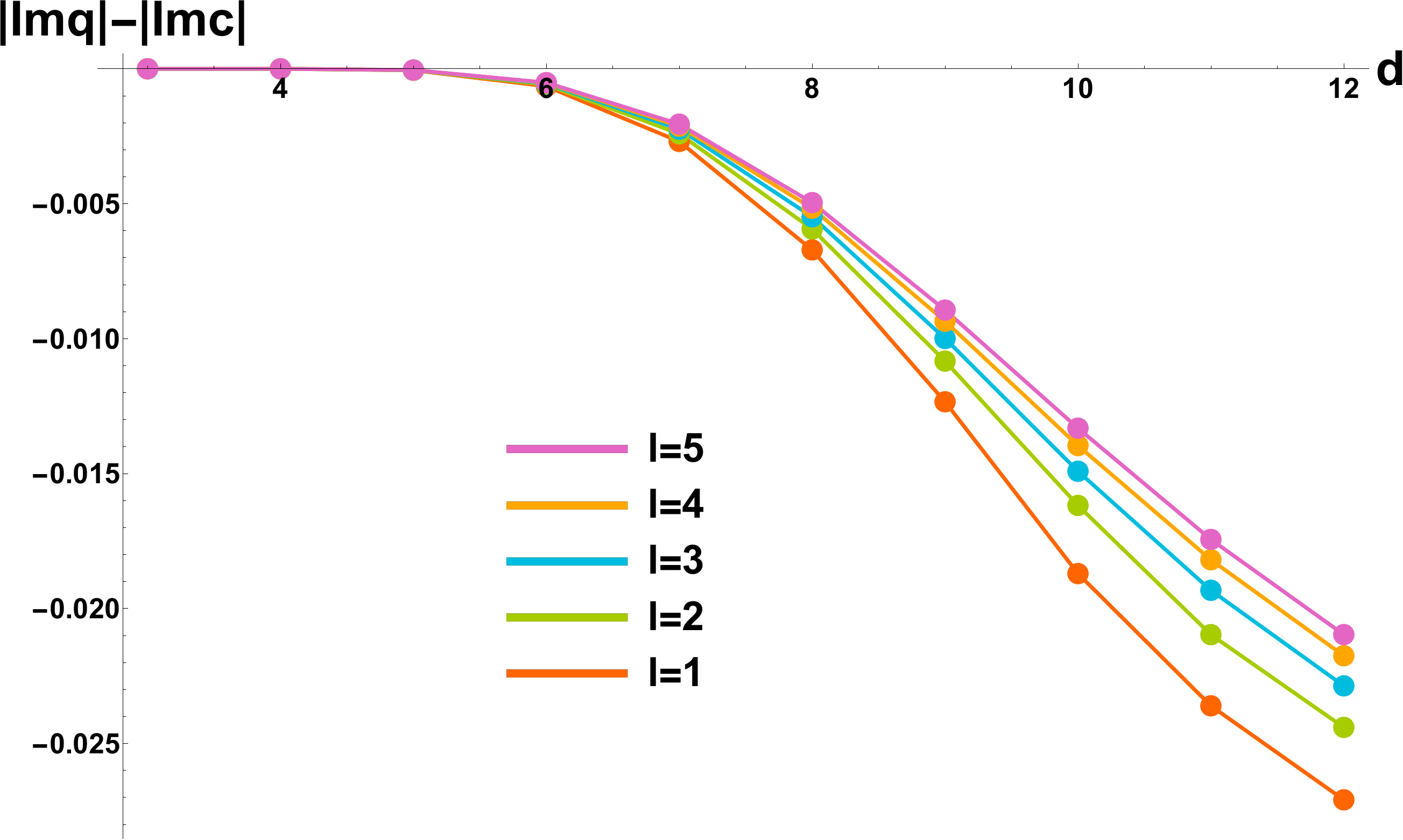}
		\caption{The difference in the absolute value of the imaginary parts of QNMs with respect to 
		the dimensions of spacetime when $M=7000$, $n=0$.}
		\label{DIm-vs-d}
	\end{figure}

	We can further investigate the influence of the dimension by Fig. \ref{DRe-vs-d} and Fig. \ref{DIm-vs-d}. 
	We observe that in various dimensions of spacetime, 
	quantum corrections consistently reduce the absolute value of the imaginary part, 
	indicating that the scalar field always decay more slowly. 
	As for the real part, when the spacetime dimensions increase, 
	the quantum corrections initially causes the real part to increase and then decrease. 
	This suggests that quantum corrections accelerate the oscillation of scalar fields in low-dimensional spacetime while 
	decelerating the oscillation of scalar fields in high-dimensional spacetime. 


\begin{figure}[!htbp]
		\includegraphics [width=0.48\textwidth]{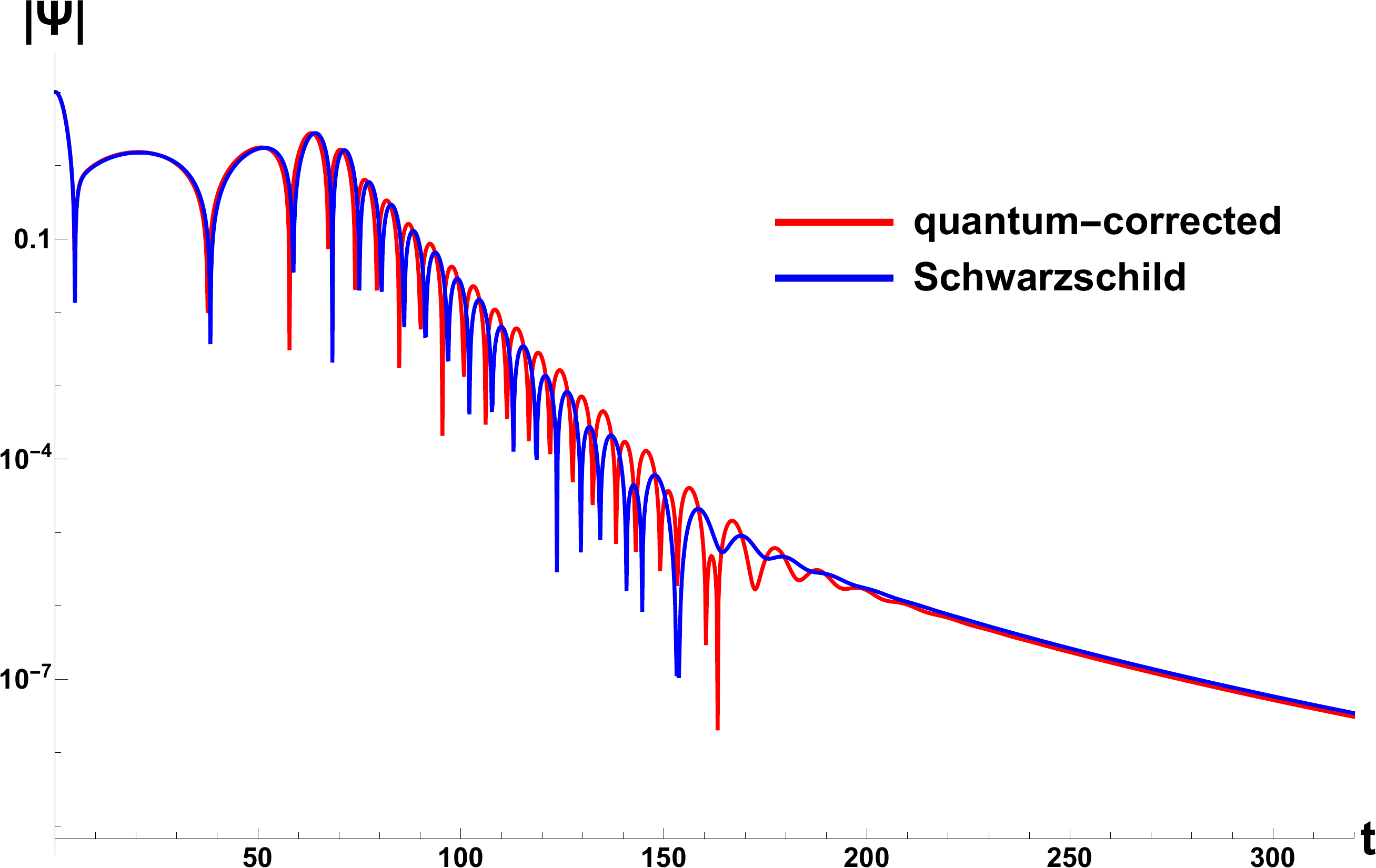}
		\caption{The evolution diagram of scalar field perturbations $\Psi$ for $d=4$, $G=1$, $M=8$, $\alpha=1$, $n=0$ and $l=2$}
		\label{time-domain-d4}
	\end{figure}

	Fig. \ref{time-domain-d4} illustrates the ringdown waveforms of scalar fields obtained by FEM in quantum-corrected and 
	Schwarzschild BHs for $d=4$, $M=8$, $n=0$ and $l=2$. 
	It can be observed that the two waveforms are nearly identical, exhibiting power-law tails. 
	This similarity arises from the similar asymptotic behavior of the effective potentials of the two BHs. 
	However, the quantum-corrected BH exhibits higher oscillation frequency and 
	lower damping rate compared to the Schwarzschild BH. 
	This observation aligns with the conclusions drawn from the WKB approximation. 

	\section{Thermodynamics}\label{thermodynamics}
	We are interested in the thermodynamic properties of quantum-corrected BHs. 
	We will mainly focus on the five-dimensional case and calculate the Hawking temperature, 
	entropy, heat capacity and free energy of the BH. 
	Then we will compare the results to those of other dimensions to highlight the effects of the dimensions. 
	We set the parameter as $G\to1$.

	\subsection{Hawking Temperature}
\begin{figure}[!htbp]
		\includegraphics [width=0.48\textwidth]{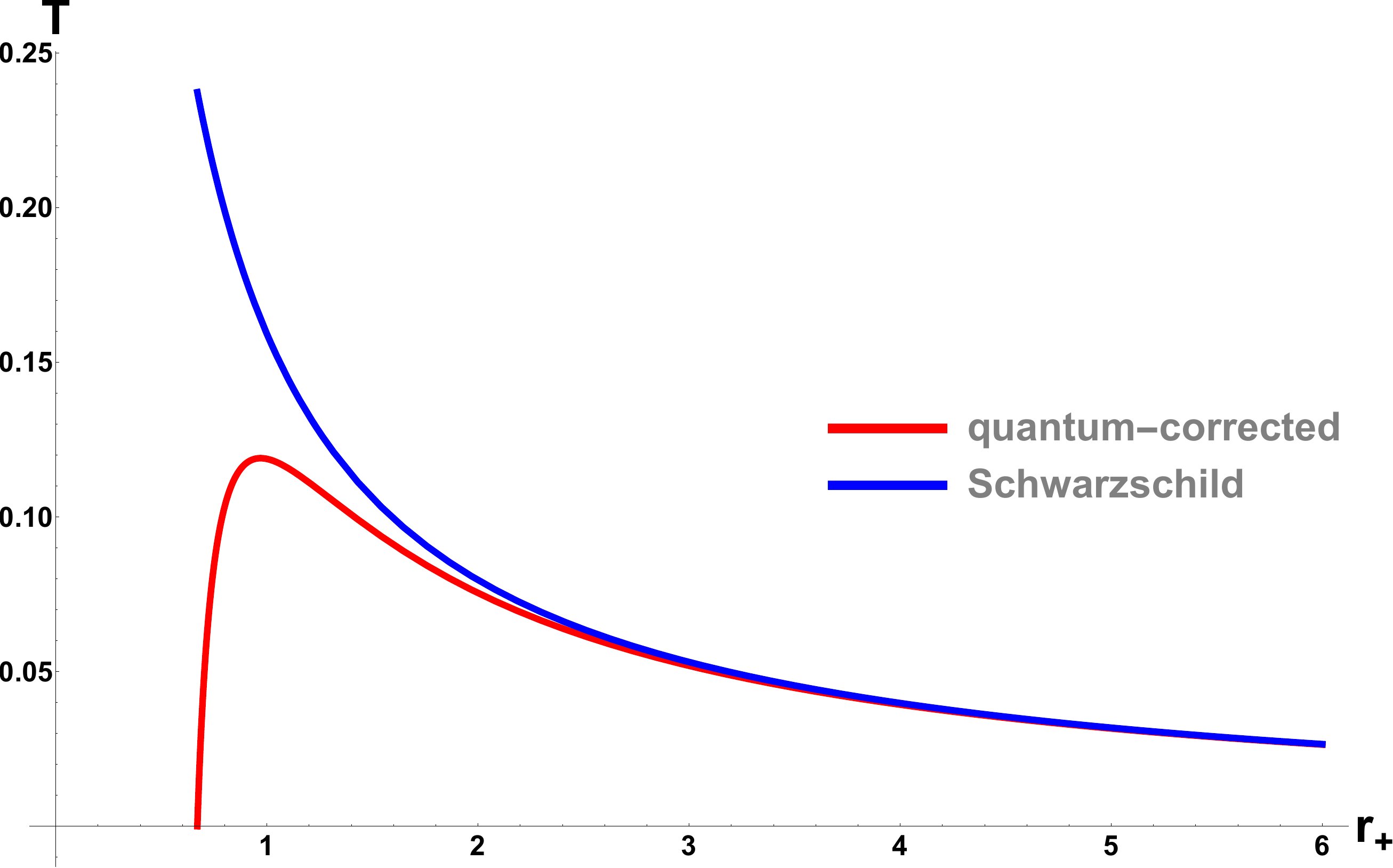}
		\caption{Hawking temperature of the quantum-corrected and Schwarzschild black hole when $d=4$, $G=1$, $\alpha=0.1$.}
		\label{T-vs-rh-4}
	\end{figure}

	For the event horizon, $f(r_{+})=0$, the mass $M$ of the BH can be expressed as a function of $r_{+}$. 
	In the five-dimensional quantum-corrected case, one has 
	\begin{equation}\label{M5}
		M_{d=4}=\frac{3\pi r_{+}^3\left(r_{+}-\sqrt{r_{+}^2-4\alpha}\right)}{16\alpha} .
	\end{equation}
	There are multiple methods available for calculating the Hawking temperature. 
	While Hawking initially computed it by the quantum field theory on the BH background \cite{Hawking:1974rv}, 
	another approach is to assume the smoothness of the corresponding Euclidean spacetime \cite{Cai:2001dz,Cai:2003gr}. 
	The conical singularity is avoided at the event horizon in the Euclidean sector of the BH by the following definition, 
	\begin{equation}
		T=\frac{f'(r_+)}{4\pi} .
	\end{equation}
	For the five-dimensional quantum-corrected BH, the Hawking temperature is 
	\begin{equation}\label{T5}
		T_{d=4}=\frac{-r_{+}^2+r_{+}\sqrt{r_{+}^2-4\alpha}+3\alpha}{2\pi r_{+}\alpha} .
	\end{equation}
	In Fig. \ref{T-vs-rh-4}, we plot the curve of the Hawking temperature 
	with respect to $r_+$ and compare it with the classical Schwarzschild case. 
	It can be observed that for large $r_+$, 
	the Hawking temperature of a quantum-corrected BH closely resembles the classical case. 
	However, significant discrepancies arise for small $r_+$. 
	The Hawking temperature of a Schwarzschild BH diverges towards infinity, 
	while that of a quantum-corrected BH first increases to 
	a peak and then rapidly drops to zero as the $r_+$ decreases. 
	Moreover, the Hawking temperature of the quantum-corrected BH 
	reaches zero for the extreme case of $r_{+}=r_{-}\equiv r_e$. 
	There are similar phenomena in spacetime with other dimensions. 

	\subsection{Black Hole Entropy}
\begin{figure}[!htbp]
		\includegraphics [width=0.48\textwidth]{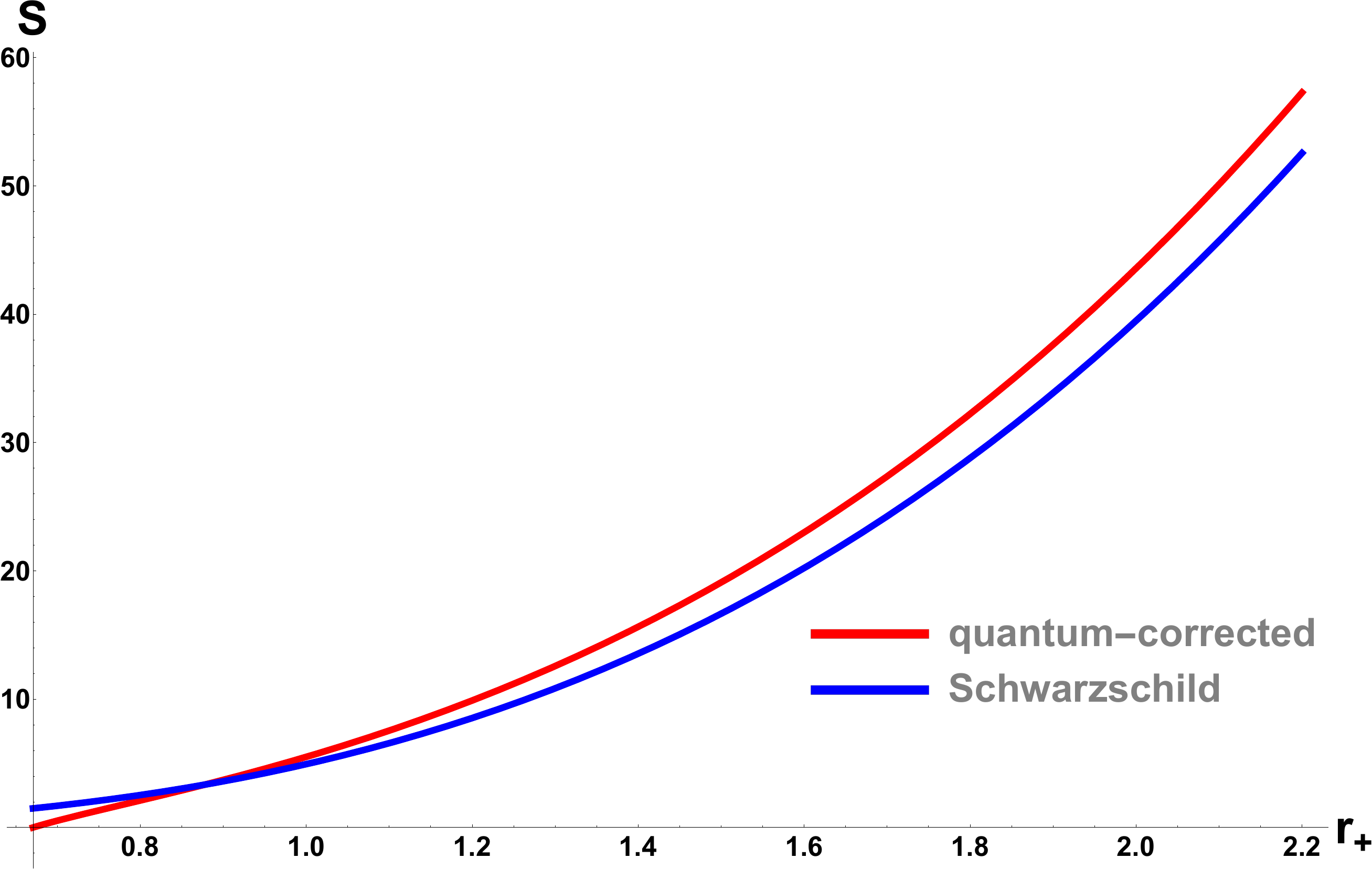}
		\caption{Entropy of the quantum-corrected and Schwarzschild black hole when $d=4$, $G=1$, $\alpha=0.1$.}
		\label{S-vs-rh-4}
	\end{figure}

	We postulate that BHs behave as thermodynamic systems and therefore 
	satisfy the first law of thermodynamics, $dM=TdS$ \cite{Cai:2001dz,Cai:2003gr}.
	Thus, we obtain 
	\begin{equation}\label{Sformula}
		S=\int T^{-1}dM=\int_{r_{e}}^{r_+}T^{-1}\left(\frac{d M}{d r}\right)dr .
	\end{equation}
	Note that the lower limit of the integral in Eq. \eqref{Sformula} cannot be zero for this particular model. 
	The integration commences from $r=r_e$ as we assume that the entropy of an extremal BH is zero. 
	In Fig. \ref{S-vs-rh-4}, we plot the curve of the BH entropy with respect to $r_+$ and 
	compare it with that of the Schwarzschild case. 
	In the case of a very small $r_+$, 
	it can be observed that the entropy of a quantized BH is smaller than that of Schwarzschild. 
	However, for other scenarios, quantum corrections consistently lead to an increase of the BH entropy. 
	We also observe similar phenomena in other dimensions of spacetime. 
	
	The entropy of the quantum-corrected BH resulted from Eq. \eqref{Sformula} reads 
	\begin{eqnarray}\label{S5}
		S_{d=4}&&=\frac{1}{2}\pi^{2}r_{+}^{2}\sqrt{r_{+}^{2}-4\alpha}+4\pi^{2}\sqrt{r_{+}^{2}-4\alpha}\alpha-\frac{25\pi^{2}\alpha^{3/2}}{4\sqrt{2}}   \notag \\
			   &&=\frac{\pi^2r_{+}^3}2+3\pi^2r_{+}\alpha+O(\frac{1}{r_{+}}) .
	\end{eqnarray}
	As $\alpha$ is a very small quantity, 
	we can employ a Taylor expansion here and thoroughly neglect higher-order terms. 
	In Eq. \eqref{S5}, the first term represents the entropy as determined by the area formula, 
	while the second term is considered as a quantum correction. 
	In five-dimensional spacetime, the quantum correction term is directly 
	proportional to the event horizon radius. 
	Similarly, we can derive expressions for other dimensional quantum BHs, e.g., 
	\begin{eqnarray}\label{S4}
		S_{d=3}&&=\pi r_{+}^2+2\pi\alpha\mathrm{~Log}[\frac{r_{+}^2}{3\alpha}]+O(\frac{1}{r_{+}}) ,
	\end{eqnarray}
	\begin{eqnarray}\label{S6}
		S_{d=5}&&=\frac{2\pi^{2}r_{+}^{4}}{3}+\frac{8}{3}\pi^{2}r_{+}^{2}\alpha +O(\frac{1}{r_{+}}) .
	\end{eqnarray}
	We find that the entropy of a quantum-corrected BH can always be 
	decomposed into the Beikenstein-Hawking area formula and a quantum correction term. 
	Particularly, we observe that a logarithmic term appeared as the leading order correction 
	to the Beikenstein-Hawking entropy in four dimensions.
	The relationship between spacetime dimensions and 
	the correction terms is summarized in Table \ref{entropy-corrections}.  
	\begin{table}[!htb]
		\caption{Relationship between spacetime dimensions and correction terms in the entropy of quantum-corrected BHs.}
		\begin{ruledtabular}
			\begin{tabular}{cc}
				$d$			&	the correction term is positively correlated with			\\
				\hline
				3			&	$\mathrm{~Log}[r^2_{+}]$										\\
				4			&	$r_+$														\\
				5			&	$r_+^2$														\\
				6			&	$r_+^3$														\\
				7			&	$r_+^4$														\\
				$\cdots$	&	$\cdots$													\\
			\end{tabular}
		\end{ruledtabular}
		\label{entropy-corrections}
	\end{table}

	\subsection{Heat Capacity}
\begin{figure}[!htbp]
		\includegraphics [width=0.48\textwidth]{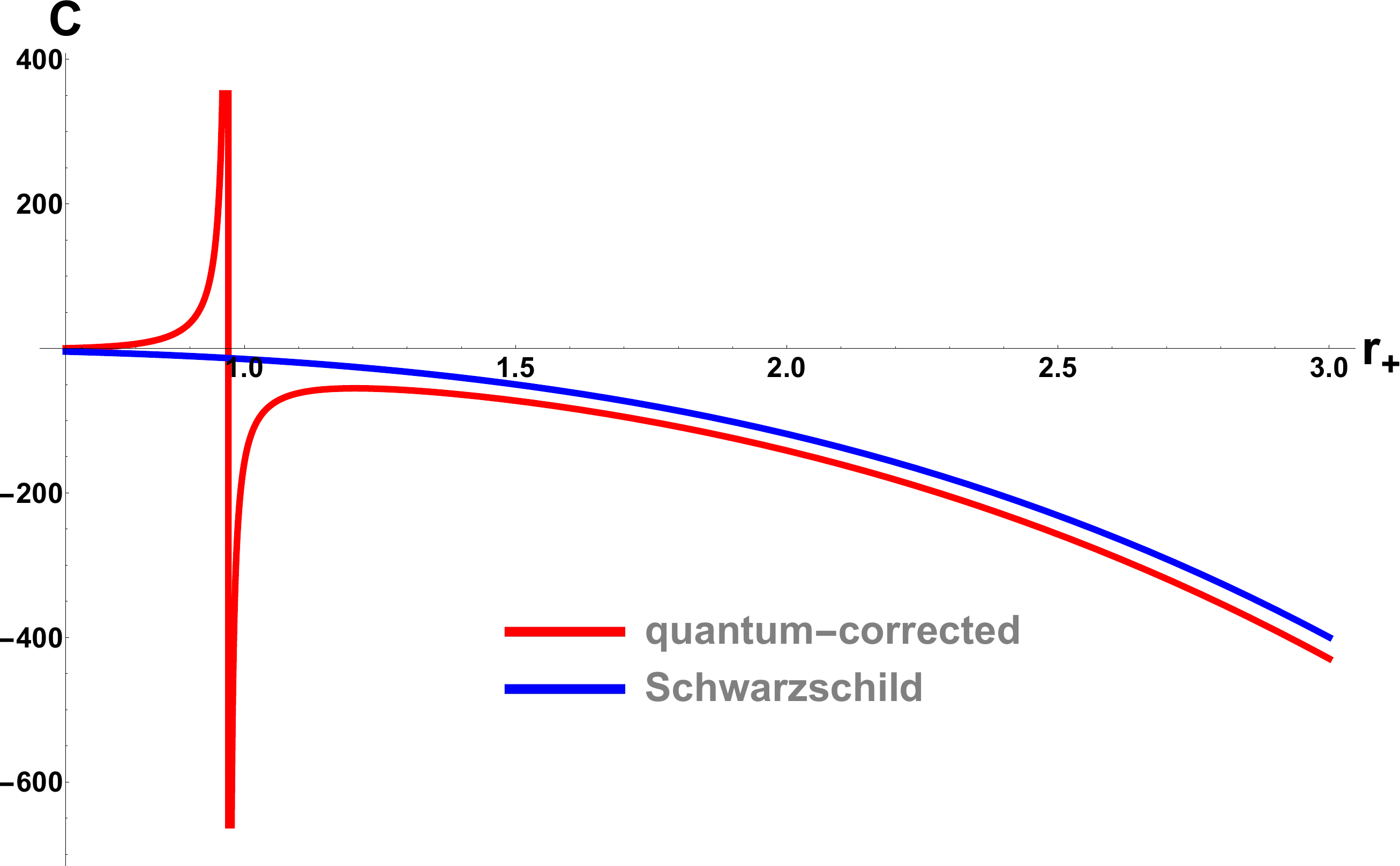}
		\caption{Heat capacity of the quantum-corrected and Schwarzschild black hole when $d=4$, $G=1$, $\alpha=0.1$.}
		\label{C-vs-rh-4}
	\end{figure}

	In BH physics, the heat capacity serves as a crucial indicator for the stability against the Hawking radiation. 
	It can be determined through the following equation \cite{Cai:2001dz,Cai:2003gr,Balart:2024rts,Chatzifotis:2023ioc}, 
	\begin{equation}
		C=T\left(\frac{d S}{d T}\right)=T\frac{\left(\frac{d S}{d r_+}\right)}{\left(\frac{d T}{d r_+}\right)} .
	\end{equation}
	Thus, the heat capacity of the five-dimensional quantum-corrected BH is 
	\begin{equation}
		C_{d=4}=\frac{3\pi^2r_+^4\left(r_+^2-r_+\sqrt{r_+^2-4\alpha}-3\alpha\right)}{-2r_+^3+2r_+^2\sqrt{r_+^2-4\alpha}+6\sqrt{r_+^2-4\alpha}\alpha} .
	\end{equation}
	In Fig. \ref{C-vs-rh-4}, we plot the curve of the heat capacity with respect to $r_+$ and compare it with that of the classical Schwarzschild case. 
	As can be seen, for large $r_+$, 
	the heat capacity of a quantum-corrected BH is in close proximity to that of the classical scenario. 
	However, the heat capacity of the quantum-corrected BH exhibits a discontinuity when $r_+$ is small. 
	This phenomenon indicates that the quantum correction introduces an extra phase transition. 
	Similar phenomena can be observed in other dimensions of spacetime as well.

	\subsection{Free Energy}
\begin{figure}[!htbp]
		\includegraphics [width=0.48\textwidth]{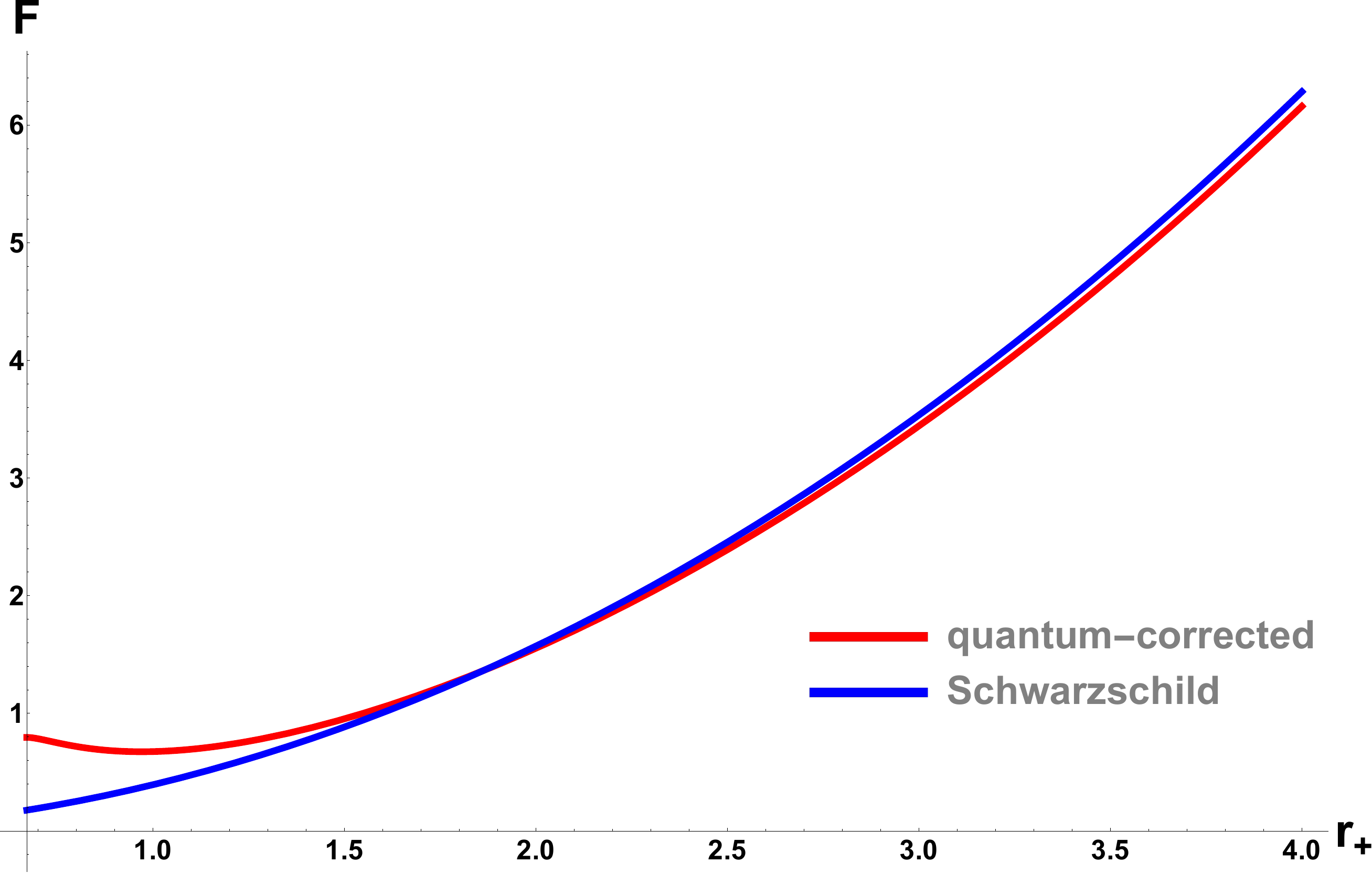}
		\caption{Free energy of the quantum-corrected and Schwarzschild black hole when $d=4$, $G=1$, $\alpha=0.1$.}
		\label{F-vs-rh-4}
	\end{figure}

	The free energy of BHs is defined as 
	\begin{equation}\label{Fdefination}
		F=M-TS .
	\end{equation}
	By substituting Eqs. \eqref{M5}, \eqref{T5}, \eqref{S5} into Eq. \eqref{Fdefination}, 
	we can derive the $F(r_+)$ relation for the five-dimensional quantum-corrected BH. 
	We plot the $F(r_+)$ curve in Fig. \ref{F-vs-rh-4} and compare it with that of the classical Schwarzschild case. 
	For small $r_+$, 
	it can be seen that the free energy of a quantized BH is greater than that of Schwarzschild. 
	However, for other scenarios, quantum corrections consistently lead to a decrease of free energy. 
	We also observe similar phenomena in other dimensions of spacetime. 

	\section{Conclusion}\label{conclusion}
	In previous sections, We have studied the quantum Oppenheimer-Snyder model in higher-dimensional spacetimes and obtain effective higher dimensional quantum-corrected BHs \eqref{QES} for the first time. 
	The interior of a collapsing star is described by a $(d+1)$-dimensional loop quantum cosmology model. 
	The exterior spacetime of the star is assumed to be static and spherically symmetric. 
	According to the Darmois-Israel junction conditions, 
	the metrics and extrinsic curvatures of the interior and exterior spacetimes should match at the stellar surface. 
	Thus, the exterior spacetime metric is uniquely determined, 
	presumed to encapsulate the correction from loop quantum cosmology. 

	Through a zero-point analysis of the quantum-corrected metric, 
	we find a lower bound on the mass of BHs formed by gravitational collapse. 
	Additionally, we perform a maximal extension of the spacetime. 
	For cases where the stellar mass is sufficient to form a BH, 
	a quantum bounce appears as the star collapses to its minimal radius, initiating an expansion. 
	This prevents the occurrence of collapse singularities and 
	leads to phenomena where BHs convert into white holes. 
	Quantum bounces also occur for cases where the stellar mass is insufficient to form a BH. 

	In order to explore its physical properties, we further compute the massless scalar perturbations in higher-dimensional quantum-corrected BH configuration. 
	We observe that larger BH masses result in lower oscillation frequencies and damping rates for the scalar waves. 
	Furthermore, for BHs of a certain mass, 
	higher spacetime dimensions lead to higher oscillation frequencies and damping rates for the scalar waves. 
	We then investigate changes induced by quantum corrections compared to the Schwarzschild case. 
	We find that in low-dimensional spacetimes, quantum corrections consistently increase oscillation frequencies, 
	whereas in high-dimensional spacetimes, they consistently decrease oscillation frequencies. 
	However, quantum corrections consistently reduce damping rates across different dimensions. 

	Moreover, the thermodynamic properties of quantum-corrected BHs are also investigated. 
	We find that the Hawking temperature of quantum-corrected BHs decreases as the event horizon radius decreases, which is opposite to the classical situation. 
	The entropy of quantum-corrected BHs in different dimensions has also been studied, 
	and it can always be divided into the Beikenstein-Hawking area formula part and a quantum correction term. 
	The precise form of the correction term depends on the dimension of spacetime. Particularly, in four-dimensions, 
	we observe that a logarithmic term appears as the leading order correction to the Beikenstein-Hawking entropy, 
	which is consistent with the calculation from full LQG \cite{Meissner:2004ju}. 
	Moreover, the calculation of the heat capacity shows that the quantum correction introduces an extra phase transition.

	\begin{acknowledgments}
		We acknowledges the valuable discussions with Dr. Cong Zhang. 
		This work is supported by National Natural Science Foundation of China (NSFC) with Grants No.12275087 and No.12275022.
	\end{acknowledgments}

	\appendix

	\section{The junction conditions}\label{Append1}
	We consider the interior of the star described by the following metric, 
	\begin{equation}\label{dsin}
		\mathrm{d}s_{in}^2=-\mathrm{d}T^2+a^2(T)[\mathrm{d}R^2+R^2\mathrm{d}\Omega^2] .
	\end{equation}
	The surface of the star corresponds to the radial coordinate $R_0$.
	Therefore, it is a $d$-dimensional hypersurface 
	induced from Eq. \eqref{dsin}, which can be described in the following metric, 
	\begin{equation}
		\left.\mathrm{d}s_{in}^2\right|_\Sigma=-\mathrm{d}T^2+a^2(T)R_0^2\mathrm{d}\Omega^2 .
	\end{equation}
	At the same time, we want to induced this hypersurface from the exterior spacetime which is described by 
	\begin{equation}
		\mathrm{d}s_{out}^2=-f(r)\mathrm{d}t^2+g(r)^{-1}\mathrm{d}r^2+r^2\mathrm{d}\Omega^2 ,
	\end{equation}
	so that we can match the metrics and make sure the surface is unique.
	It is worth noting that, according to our assumption, 
	every point on the surface of the star always corresponds to the same proper time $T$ 
	whose value will change as the collapse progresses. 
	Therefore, when considering the metric of the hypersurface from the exterior, 
	it is convenient to do the parameterization with $T$. 
	Then, we get 
	\begin{equation}
		\left.\mathrm{d}s_{out}^2\right|_\Sigma=-(f\dot{t}^2-g^{-1}\dot{r}^2)\mathrm{d}T^2+r^2(T)\mathrm{d}\Omega^2 ,
	\end{equation}
	where $r$ and $t$ are both $T$-dependent. And $\dot{t}$ and $\dot{r}$ are derivatives with respect to $T$. 
	According to the Darmois-Israel junction conditions, 
	the first and second fundamental forms of the induced metric obtained from the interior and exterior must be equal.
	The equality of the first fundamental form implies that the metric components are equal, which leads to 
	\begin{equation}\label{g00}
		1=f\dot{t}^2-g^{-1}\dot{r}^2
	\end{equation}
	and 
	\begin{equation}\label{g11}
		a(T)R_0=r(T) .
	\end{equation}
	The equality of the second fundamental form implies that the extrinsic curvature is equal. 
	From the interior spacetime, we can easily get 
	\begin{equation}\label{KTTin}
		K_{TT}^{-}=0 
	\end{equation}
	and 
	\begin{equation}\label{K11in}
		K_{\theta_1\theta_1}^{-}=\frac{K_{\theta_2\theta_2}^{-}}{\sin^{2}\theta_1}=\frac{K_{\theta_3\theta_3}^{-}}{\sin^{2}\theta_1\sin^{2}\theta_2}=\cdots=a(T)R_0 .
	\end{equation}
	However, when considering it from the exterior spacetime, we have better introduce a conserved quantity. 
	The exterior metric is assumed to be static so that $\partial/\partial t$ is a Killing vector. 
	Then we get
	\begin{equation}
		-\mathrm{g}_{ab}^{out}\left(\frac{\partial}{\partial t}\right)^a\left(\frac{\partial}{\partial T}\right)^b=f\dot{t}\equiv E ,
	\end{equation}
	which is the conserved energy during the collapse process. 
	We can first obtain $K_{ab}^{+}$ in the Schwarzschild coordinates whose non-zero components are 
	\begin{equation}\label{K11out}
		K_{\theta_1\theta_1}^{+}=\frac{K_{\theta_2\theta_2}^{+}}{\sin^{2}\theta_1}=\frac{K_{\theta_3\theta_3}^{+}}{\sin^{2}\theta_1\sin^{2}\theta_2}=\cdots=rE\sqrt{f^{-1}g} .
	\end{equation}
	It means that 
	\begin{eqnarray}
		K_{TT}^{+}&&=K_{ab}^{+}\left(\frac{\partial}{\partial T}\right)^{a}\left(\frac{\partial}{\partial T}\right)^{b}   \notag \\
				  &&=K_{ab}^{+}\left[\dot{t}\left(\frac{\partial}{\partial t}\right)^{a}+\dot{r}\left(\frac{\partial}{\partial r}\right)^{a}\right]\left[\dot{t}\left(\frac{\partial}{\partial t}\right)^{b}+\dot{r}\left(\frac{\partial}{\partial r}\right)^{b}\right]   \notag \\
				  &&=\dot{t}^{2}K_{tt}^{+}+2\dot{t}\dot{r}K_{tr}^{+}+\dot{r}^{2}K_{rr}^{+}   \notag \\
				  &&=0 ,
	\end{eqnarray}
	which agrees with Eq. \eqref{KTTin}.
	By comparing Eq. \eqref{K11in} with Eq. \eqref{K11out}, we get 
	\begin{equation}\label{K11}
		a(T)R_0=rE\sqrt{f^{-1}g} .
	\end{equation}
	After combining Eq. \eqref{K11} with Eq. \eqref{g00} and Eq. \eqref{g11}, we obtain 
	\begin{equation}
		f=E^{2}g
	\end{equation}
	and 
	\begin{equation}\label{gofrdot}
		g=1-\dot{r}^2 .
	\end{equation}
	By utilizing the definition of the Hubble parameter, 
	Eq. \eqref{gofrdot} and the exterior metric can be expressed respectively as 
	\begin{equation}
		g=1-H^2r^2
	\end{equation}
	and 
	\begin{equation}
		\mathrm{d}s_{out}^2=-(1-H^2r^2)E^{2}\mathrm{d}t^2+(1-H^2r^2)^{-1}\mathrm{d}r^2+r^2\mathrm{d}\Omega^2 .
	\end{equation}
	To simplify without sacrificing generality, we can select $E=1$ and it leads to $f=g$, 
	which can be interpreted as performing a coordinate transformation $t\to t/E$. 
	In conclusion, the expression for the exterior metric can be simplified as 
	\begin{equation}\label{dsout}
		\mathrm{d}s_{out}^2=-(1-H^2r^2)\mathrm{d}t^2+(1-H^2r^2)^{-1}\mathrm{d}r^2+r^2\mathrm{d}\Omega^2 .
	\end{equation}

	\section{Zeros of \texorpdfstring{$f(r)$}{f(r)}}\label{Append2}
\begin{figure}[!htb]
		\includegraphics [width=0.48\textwidth]{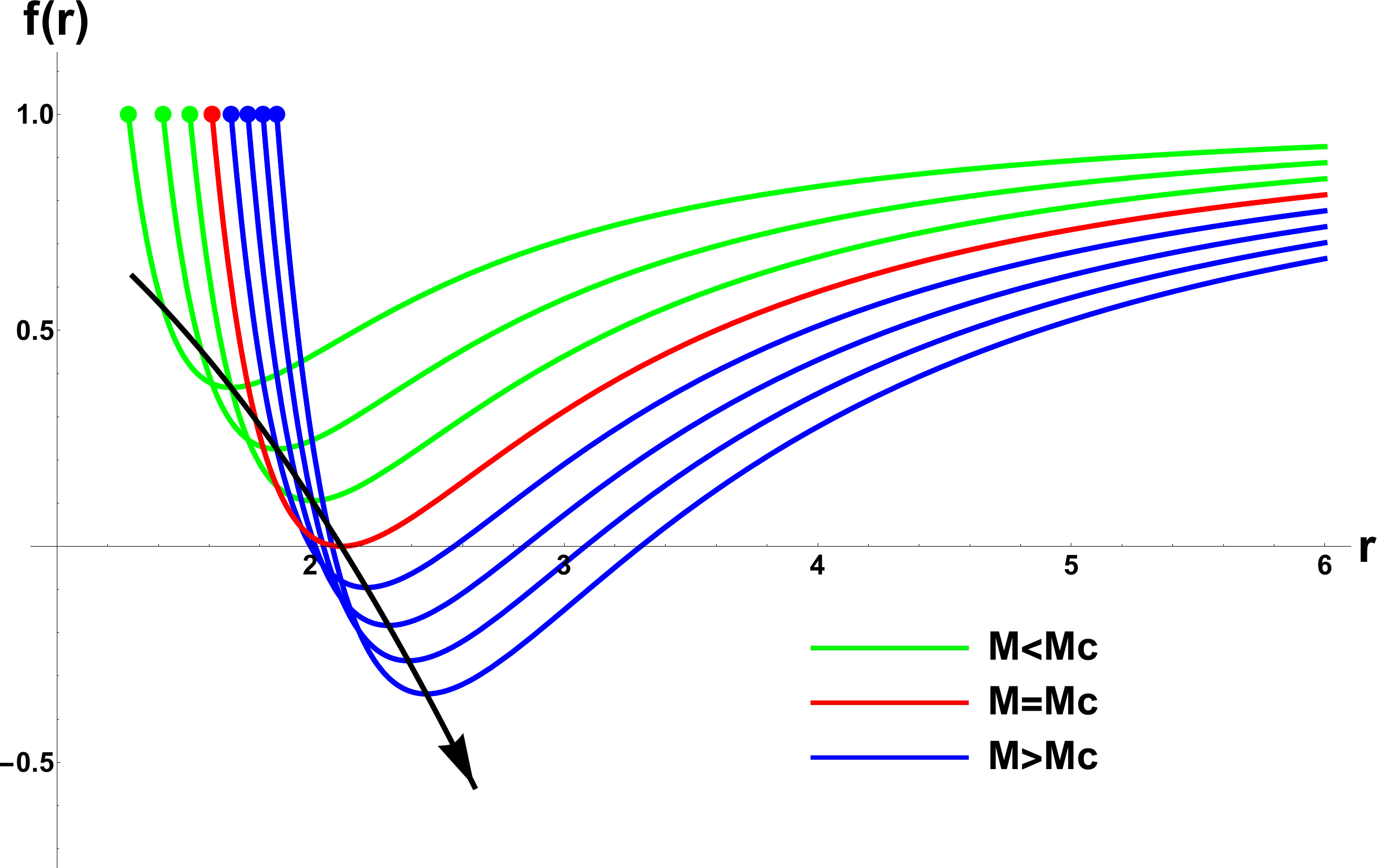}
		\caption{$f(r)$ curves for different $M$. The red line corresponds to $M=M_c$. 
		The black line with arrowhead connect the lowest points of each $f(r)$ curve, 
		indicating the trend of movement of the lowest point as $M$ increases. 
		The filled circles at the left end of each $f(r)$ curve represent the minimum radius, $r_b$, which is the bounce radius.}
		\label{fr-vs-M}
	\end{figure}

	We will discuss the function 
	\begin{equation}
		f(r)=1-\frac{2GM\mu}{r^{d-2}}+\frac{4{G}^2{M}^2\mu^2\alpha}{r^{2d-2}},\left.r\in\left[\begin{matrix}r_b,\infty\end{matrix}\right.\right)
	\end{equation}
	in an arbitrary $(d+1)$-dimensional spacetime $(d\geqslant3)$. 
	Consequently, providing an analytical expression for the zero points of 
	these numerous functions is a quite challenging task. 
	Nevertheless, we can qualitatively explore the distribution of their zero points. 
	We find that $f(r_\mathrm{b})=1$ and when $r\rightarrow\infty$, $f(r)\rightarrow1$. 
	By taking the derivative of $f(r)$ with respect to $r$ within its domain, we get
	\begin{equation}
		f'(r)=\frac{2GM\mu(d-2)}{r^{d-1}}-\frac{8G^2M^2\mu^2\alpha(d-1)}{r^{2d-1}} .
	\end{equation}
	We observe that there exists only one critical point that makes $f'(r)=0$, which is denoted as
	\begin{equation}
		r_{m}\equiv\left(\frac{4GM\mu\alpha(d-1)}{d-2}\right)^{\frac1d} .
	\end{equation}
	Furthermore, since $f(r)$ is differentiable throughout its domain and $f'(r_\mathrm{b})=-(2GM\mu)^{\frac1d}\alpha^{\frac{1-d}d}d<0$, 
	it indicates that this critical point is not only a local minimum point, but also a global minimum point.
	We proceed to determine how the minimum value, which is 
	\begin{equation}
		f(r_{m})=1-(GM\mu)^{\frac2d}(\frac{d-2}{4\alpha})^{\frac{d-2}d}(d-1)^{\frac{2-2d}d}d ,
	\end{equation}
	varies with $M$. 
	Differentiating $f(r_{m})$ with respect to $M$, we get
	\begin{equation}
		\frac{\partial f(r_{m})}{\partial M}=-2^{\frac{4-d}d}(G\mu)^{\frac2d}(M\alpha)^{\frac{2-d}d}(d-2)^{\frac{d-2}d}(d-1)^{\frac{2-2d}d} ,
	\end{equation}
	which is always smaller than zero for $\left.M\in\left(\begin{matrix}0,\infty\end{matrix}\right.\right)$. 
	We also observe that $\left.f(r_{m})\right|_{M=0}=1$, whereas as $M\rightarrow\infty$, $f(r_{m})\rightarrow{-\infty}$. 
	It means that if $M$ changes from $0$ to $\infty$, 
	the minimum value of $f(r)$ will continuously decreases from $1$ to $-\infty$. 
	In conclusion, the higher the stellar mass we consider, the more zero points $f(r)$ will have: 
	from 0 to 1, and ultimately to 2, as is shown in Fig. \ref{fr-vs-M}. 
	
	That the function $f(r)$ has exactly one zero point is a critical scenario, which is equivalent to 
	\begin{equation}\label{frme0}
		f(r_{m})=0 .
	\end{equation}
	By solving Eq. \eqref{frme0}, we get the critical mass 
	\begin{equation}
		M=(G\mu)^{-1}\left(\frac{d-2}{4\alpha}\right)^{\frac{2-d}2}(\mathrm{d-1})^{\frac{2d-2}2}d^{-\frac d2}\equiv M_c .
	\end{equation}
	The discussion holds true for any dimension.

	\section{QNMs data}\label{Append3}
	\begin{table*}[!htb]
		\caption{QNMs for quantum-corrected black holes when $n=0$, $l=1$.}
		\begin{ruledtabular}
			\begin{tabular}{ccccc}
				& {M=7000} & {M=8000} & {M=9000} & {M=10000} \\
				\hline
				{d=3}  & {0.000041848}-{0.000013951} i & {0.000036617}-{0.000012208} i & {0.000032548}-{0.000010851} i & {0.000029294}-{0.000009766} i \\
				{d=4}  & {0.013181123}-{0.004700155} i & {0.012329775}-{0.004396633} i & {0.011624597}-{0.004145215} i & {0.011028041}-{0.003932519} i \\
				{d=5}  & {0.096780315}-{0.034007083} i & {0.092564339}-{0.032531195} i & {0.088998477}-{0.031282308} i & {0.085925394}-{0.030205599} i \\
				{d=6}  & {0.272840676}-{0.092231305} i & {0.263867302}-{0.089245106} i & {0.256198226}-{0.086689101} i & {0.249527599}-{0.084463073} i \\
				{d=7}  & {0.520519866}-{0.167775301} i & {0.506778607}-{0.163500463} i & {0.494959393}-{0.159813236} i & {0.484620617}-{0.156580229} i \\
				{d=8}  & {0.813337509}-{0.249486920} i & {0.795456492}-{0.244307197} i & {0.780005907}-{0.239814837} i & {0.766435902}-{0.235856749} i \\
				{d=9}  & {1.130960711}-{0.330822226} i & {1.109741082}-{0.325064289} i & {1.091336769}-{0.320049653} i & {1.075120720}-{0.315615611} i \\
				{d=10} & {1.459571406}-{0.408825014} i & {1.435902561}-{0.402636319} i & {1.415282636}-{0.397278182} i & {1.397059274}-{0.392534929} i \\
				{d=11} & {1.791551463}-{0.483797298} i & {1.765754733}-{0.477017556} i & {1.743308980}-{0.471145484} i & {1.723466797}-{0.465976734} i \\
				{d=12} & {2.123475996}-{0.554893990} i & {2.095840384}-{0.547795969} i & {2.071776249}-{0.541608134} i & {2.050496373}-{0.536134508} i \\
			\end{tabular}
		\end{ruledtabular}
		\label{qnm-vs-M}
	\end{table*}

	\begin{table*}[!htb]
		\caption{QNMs for $M=7000$, $n=0$, $l=1$.}
		\begin{ruledtabular}
			\begin{tabular}{cccc}
				& {quantum-corrected} & {Schwarzschild} & {difference} \\
				\hline
				{d=3}  & {0.000041848}-{0.000013951} i & {0.000041848}-{0.000013951} i & ${7.056564399}\times 10^{-14}+\left({4.628174132}\times 10^{-14}\right)$ i \\
				{d=4}  & {0.013181123}-{0.004700155} i & {0.013180814}-{0.004700505} i & ${3.089909130}\times 10^{-7}+\left({3.499749140}\times 10^{-7}\right)$ i \\
				{d=5}  & {0.096780315}-{0.034007083} i & {0.096746999}-{0.034062935} i & {0.000033316}+{0.000055852} i \\
				{d=6}  & {0.272840676}-{0.092231305} i & {0.272578405}-{0.092880408} i & {0.000262271}+{0.000649104} i \\
				{d=7}  & {0.520519866}-{0.167775301} i & {0.519843789}-{0.170468073} i & {0.000676077}+{0.002692772} i \\
				{d=8}  & {0.813337509}-{0.249486920} i & {0.812656547}-{0.256199649} i & {0.000680962}+{0.006712729} i \\
				{d=9}  & {1.130960711}-{0.330822226} i & {1.131469810}-{0.343162495} i &  -0.000509099+{0.012340269} i \\
				{d=10} & {1.459571406}-{0.408825014} i & {1.463317794}-{0.427532577} i &  -0.003746387+{0.018707563} i \\
				{d=11} & {1.791551463}-{0.483797298} i & {1.799959685}-{0.507404915} i &  -0.008408222+{0.023607616} i \\
				{d=12} & {2.123475996}-{0.554893990} i & {2.136264393}-{0.581984557} i &  -0.012788396+{0.027090567} i \\
			\end{tabular}
		\end{ruledtabular}
		\label{qnm-vs-n1}
	\end{table*}

	\begin{table*}[!htb]
		\caption{QNMs for $M=7000$, $n=0$, $l=2$.}
		\begin{ruledtabular}
			\begin{tabular}{cccc}
				& {quantum-corrected} & {Schwarzschild} & {difference} \\
				\hline
				{d=3}  & {0.000069092}-{0.000013823} i & {0.000069092}-{0.000013823} i & ${1.080028265}\times 10^{-13}+\left({4.382933869}\times 10^{-14}\right)$ i \\
				{d=4}  & {0.019597066}-{0.004638019} i & {0.019596643}-{0.004638343} i & ${4.231344902}\times 10^{-7}+\left({3.241864397}\times 10^{-7}\right)$ i \\
				{d=5}  & {0.134581737}-{0.033520251} i & {0.134538773}-{0.033571424} i & {0.000042963}+{0.000051172} i \\
				{d=6}  & {0.362064961}-{0.090959254} i & {0.361737109}-{0.091546915} i & {0.000327852}+{0.000587661} i \\
				{d=7}  & {0.667192581}-{0.165712559} i & {0.666333512}-{0.168120748} i & {0.000859069}+{0.002408189} i \\
				{d=8}  & {1.015106238}-{0.246920991} i & {1.013975992}-{0.252858532} i & {0.001130246}+{0.005937541} i \\
				{d=9}  & {1.382108705}-{0.328135236} i & {1.381679188}-{0.338968680} i & {0.000429516}+{0.010833445} i \\
				{d=10} & {1.753689464}-{0.406449987} i & {1.755681350}-{0.422631002} i &  -0.001991886+{0.016181015} i \\
				{d=11} & {2.122456261}-{0.481347423} i & {2.128123023}-{0.502316672} i &  -0.005666762+{0.020969248} i \\
				{d=12} & {2.484977448}-{0.552104333} i & {2.494597601}-{0.576511127} i &  -0.009620153+{0.024406794} i \\
			\end{tabular}
		\end{ruledtabular}
		\label{qnm-vs-n2}
	\end{table*}

	\begin{table*}[!htb]
		\caption{QNMs for $M=7000$, $n=0$, $l=3$.}
		\begin{ruledtabular}
			\begin{tabular}{cccc}
				& {quantum-corrected} & {Schwarzschild} & {difference} \\
				\hline
				{d=3}  & {0.000096481}-{0.000013786} i & {0.000096481}-{0.000013786} i & ${1.481072480}\times 10^{-13}+\left({4.284383788}\times 10^{-14}\right)$ i \\
				{d=4}  & {0.026048929}-{0.004615525} i & {0.026048378}-{0.004615837} i & ${5.515596155}\times 10^{-7}+\left({3.112342058}\times 10^{-7}\right)$ i \\
				{d=5}  & {0.172553366}-{0.033317884} i & {0.172498905}-{0.033366391} i & {0.000054461}+{0.000048507} i \\
				{d=6}  & {0.451586769}-{0.090380262} i & {0.451171474}-{0.090929802} i & {0.000415295}+{0.000549540} i \\
				{d=7}  & {0.814212378}-{0.164710857} i & {0.813093902}-{0.166949210} i & {0.001118476}+{0.002238353} i \\
				{d=8}  & {1.217122531}-{0.245633197} i & {1.215498948}-{0.251116934} i & {0.001623583}+{0.005483737} i \\
				{d=9}  & {1.633280084}-{0.326737403} i & {1.632013134}-{0.336733362} i & {0.001266950}+{0.009995960} i \\
				{d=10} & {2.047563276}-{0.405131488} i & {2.048028513}-{0.420046665} i &  -0.000465237+{0.014915177} i \\
				{d=11} & {2.452683054}-{0.479838910} i & {2.455906177}-{0.499164893} i &  -0.003223123+{0.019325983} i \\
				{d=12} & {2.845533008}-{0.550374535} i & {2.851949755}-{0.573242729} i &  -0.006416748+{0.022868194} i \\
			\end{tabular}
		\end{ruledtabular}
		\label{qnm-vs-n3}
	\end{table*}

	\begin{table*}[!htb]
		\caption{QNMs for $M=7000$, $n=0$, $l=4$.}
		\begin{ruledtabular}
			\begin{tabular}{cccc}
				& {quantum-corrected} & {Schwarzschild} & {difference} \\
				\hline
				{d=3}  & {0.000123917}-{0.000013770} i & {0.000123917}-{0.000013770} i & ${1.889714784}\times 10^{-13}+\left({4.237305697}\times 10^{-14}\right)$ i \\
				{d=4}  & {0.032514884}-{0.004605036} i & {0.032514199}-{0.004605340} i & ${6.849231839}\times 10^{-7}+\left({3.042701623}\times 10^{-7}\right)$ i \\
				{d=5}  & {0.210603409}-{0.033216061} i & {0.210536826}-{0.033262978} i & {0.000066582}+{0.000046917} i \\
				{d=6}  & {0.541260779}-{0.090071997} i & {0.540753993}-{0.090598691} i & {0.000506786}+{0.000526693} i \\
				{d=7}  & {0.961420295}-{0.164160097} i & {0.960024501}-{0.166286337} i & {0.001395794}+{0.002126240} i \\
				{d=8}  & {1.419333036}-{0.244907445} i & {1.417137689}-{0.250076549} i & {0.002195347}+{0.005169104} i \\
				{d=9}  & {1.884539399}-{0.325963563} i & {1.882334320}-{0.335318903} i & {0.002205079}+{0.009355340} i \\
				{d=10} & {2.341250785}-{0.404331569} i & {2.340219926}-{0.418294109} i & {0.001030859}+{0.013962539} i\\
				{d=11} & {2.782333533}-{0.478932606} i & {2.783493569}-{0.497126850} i &  -0.001160036+{0.018194244} i \\
				{d=12} & {3.205296572}-{0.549256783} i & {3.209171962}-{0.571007908} i &  -0.003875390+{0.021751125} i \\
			\end{tabular}
		\end{ruledtabular}
		\label{qnm-vs-n4}
	\end{table*}

	\begin{table*}[!htb]
		\caption{QNMs for $M=7000$, $n=0$, $l=5$.}
		\begin{ruledtabular}
			\begin{tabular}{cccc}
				& {quantum-corrected} & {Schwarzschild} & {difference} \\
				\hline
				{d=3}  & {0.000151373}-{0.000013762} i & {0.000151373}-{0.000013762} i & ${2.301299071}\times 10^{-13}+\left({4.211756134}\times 10^{-14}\right)$ i \\
				{d=4}  & {0.038987814}-{0.004599328} i & {0.038986994}-{0.004599629} i & ${8.199537709}\times 10^{-7}+\left({3.000846118}\times 10^{-7}\right)$ i \\
				{d=5}  & {0.248695674}-{0.033158047} i & {0.248616717}-{0.033203963} i & {0.000078956}+{0.000045917} i \\
				{d=6}  & {0.631024851}-{0.089890947} i & {0.630424001}-{0.090402519} i & {0.000600849}+{0.000511573} i \\
				{d=7}  & {1.108748450}-{0.163829537} i & {1.107068528}-{0.165881023} i & {0.001679922}+{0.002051486} i \\
				{d=8}  & {1.621655328}-{0.244463417} i & {1.618884236}-{0.249422786} i & {0.002771092}+{0.004959369} i \\
				{d=9}  & {2.135886885}-{0.325467119} i & {2.132727670}-{0.334408358} i & {0.003159214}+{0.008941239} i \\
				{d=10} & {2.634871506}-{0.403821867} i & {2.632429735}-{0.417140593} i & {0.002441771}+{0.013318726} i\\
				{d=11} & {3.111745361}-{0.478316723} i & {3.111031339}-{0.495757535} i & {0.000714021}+{0.017440812} i\\
				{d=12} & {3.564639534}-{0.548488970} i & {3.566286901}-{0.569456770} i &  -0.001647367+{0.020967800} i \\
			\end{tabular}
		\end{ruledtabular}
		\label{qnm-vs-n5}
	\end{table*}

	\bibliographystyle{unsrt}
	\clearpage

\end{CJK}
\end{document}